\newif\ifHEVEA \newif\ifDRAFT \newif\ifFINAL \newif\ifREVWS \newif\ifEXPLE \newif\ifACCEPTED

\HEVEAfalse
\DRAFTfalse
\REVWSfalse
\EXPLEfalse
\FINALtrue
\ACCEPTEDtrue

\documentclass[sigconf, language=french, language=english]{acmart} 

\def\linescale{1}

\newcommand{\docinfo}{}

\ifHEVEA
\providecommand{\tabularnewline}{\\}
  \newcommand{\url}{}
  \newcommand{\hyperref}[2]{ #1}
  \newcommand{\g}[1]{ #1}

\fi

\usepackage[normalem]{ulem} 

\usepackage{xcolor}

\ifREVWS
  \newcommand{\reviews}[1]{\textcolor{orange}{#1}} \else
  \newcommand{\reviews}[1]{} \fi

\ifEXPLE
  \newcommand{\example}[1]{\textcolor{yellow}{#1}} \else
  \newcommand{\example}[1]{} \fi

\ifDRAFT
\newcommand{\barre}[1]{\textcolor{lightgray}{\sout{#1}}} \newcommand{\barrePT}[1]{\textcolor{lightgray}{\sout{#1}}} 

  \newcommand{\remark}[1]{}  

    \newcommand{\rouge}[1]{\textcolor{red}{#1}}

\else
  \newcommand{\barre}[1]{} \newcommand{\barrePT}[1]{} 

  \ifFINAL
    \renewcommand{\reviews}[1]{} \newcommand{\remark}[1]{}   \newcommand{\rouge}[1]{#1} \else
    \newcommand{\remark}[1]{\textcolor{red}{\textsc{#1}}}   \newcommand{\rouge}[1]{\textcolor{red}{#1}} \fi

         \fi

\ifDRAFT
  \usepackage{soul}
  \definecolor{bleuclair}{rgb}{0.7, 0.7, 1.0}
  \definecolor{rosepale}{rgb}{1.0, 0.7, 1.0}
  \newcommand{\hljaune}[1]{\hl{#1}}

\else
  \newcommand{\hljaune}[1]{#1}

\fi

\def\ie{{i.e.,} }

\def\cad{{c.\babelhyphen{nobreak}\`a\babelhyphen{nobreak}d.,} }

\def\pex{{p.~ex.,} }

\newcommand{\ita}[1]{\textit{#1}}

\setcitestyle{notesep={: }}

\usepackage{eurosym}

\DeclareTextSymbol{\deg}{T1}{6}
\DeclareTextSymbol{\deg}{OT1}{23}

\ifFINAL
  \newcommand{\todo}{}
\else
 \newcommand{\todo}{\rouge{TODO} \hljaune{< < < < < < < < < < < < < < < < < < <}}
\fi

\setcitestyle{nocompress}

\usepackage{hyperref}

\addto\extrasenglish{

}

\addto\extrasfrench{

}

\usepackage{subcaption}

\newcommand{\killpunct}[1]{}

\usepackage{colortbl} 

\usepackage{enumitem}

\usepackage{printlen}

\def\ie{{i.e.,} }

\def\raR{\raggedright}

\DeclareTextSymbol{\deg}{T1}{6}
\DeclareTextSymbol{\deg}{OT1}{23}

\ifHEVEA
  \providecommand{\tabularnewline}{\\}
  \newcommand{\url}{}
  \newcommand{\hyperref}[2]{ #1}
  \newcommand{\g}[1]{"#1"}
\else
  \newcommand{\g}[1]{``#1''}
\fi

\usepackage{amsmath}

\def\myshorttitle{Initiating and Replicating Interactional Properties by Optimizing Applicative Prototypes}
\def\mytitle{Initiating and Replicating the Observations of Interactional Properties by User Studies Optimizing Applicative Prototypes}

\def\mytitleFR{Générer et répliquer des observations de propriétés interactionnelles par des études expérimentales optimisant des prototypes applicatifs}

\def\mykeywords{research methodology, user-centered design, applicative cases, experimental HCI, empirical studies, user studies, user experiments, serendipity, opportunistic research, replication, results replication, replication studies, property studies, interaction loops, interaction prisms, interactional properties} 

\def\mykeywordsFR{méthodologie de recherche, conception centrée utilisateur, cas applicatifs, IHM expérimentale, études empiriques, études utilisateurs, expérimentations utilisateurs, sérendipité, recherche opportuniste, réplication, réplication de résultats, études réplicatives, études de propriétés, boucles interactionnelles, prismes interactionnels, propriétés interactionnelles}


\AtBeginDocument{}

\setcopyright{cc}
\copyrightyear{2025}
\acmYear{2025}
\acmDOI{XXXXXXX.XXXXXXX}

\acmConference[]{arXiv.org, Cornell University}{2025}{New York, USA}
\acmPrice{}
\acmISBN{}

\begin{document}

\title[\myshorttitle]{\mytitle}
\translatedtitle{french}{\mytitleFR}

\author{Guillaume Rivière}
\email{g.riviere@estia.fr}
\orcid{0000-0001-8390-9751}
\affiliation{\institution{Univ. Bordeaux, ESTIA-Institute of Technology, EstiaR}
  \city{Bidart}
  \country{France}
  \postcode{F-64210}
}
\renewcommand{\shortauthors}{Riviere}

\begin{abstract}

The science of Human--Computer Interaction (HCI) is populated by isolated empirical findings, often tied to specific technologies, designs, and tasks. This paper proposes a formalization of user interaction observations (instead of user interfaces) and an associated revealing method (interaction loop diffraction). The resulting interactional properties that are studied in a calibrated manner, are well suited to replication across various conditions (prototypes, technologies, tasks, and user profiles). In particular, interactional properties can emerge and be replicated within the workflow of applicative cases, which in return benefit from the optimization of applicative prototypes. Applicative cases' publications will then contribute to demonstrating technology utility, along with providing empirical results that will lead future work to theory consolidation and theory building, and finally to a catalog and a science of relevant interactional properties. These properties will contribute to better user interactions, especially for the variety of ubiquitous user interfaces.

\end{abstract}
 
\begin{translatedabstract}{french}

La science des Interactions Humain-Machine (IHM) revêt beaucoup de résultats empiriques isolés, souvent liés à des technologies, des conceptions et des tâches spécifiques. Le présent article propose un formalisme d'observation de l'interaction (plutôt que des interfaces) adossé à une méthode de révélation (la diffraction de boucles d'interaction). Les propriétés interactionnelles ainsi étudiées sont propices à des réplications au travers de conditions variées (prototypes, technologies, tâches et profils de personnes utilisatrices). Surtout, elles peuvent émerger et être répliquées lors du traitement de cas applicatifs, qui bénéficient en retour d'une optimisation des prototypes.
La publication de cas applicatifs continue alors d'illustrer l'utilité des technologies, tout en produisant des résultats qui conduiront de futures recherches à consolider ou établir des théories, et à aboutir à un catalogue et une science de propriétés interactionnelles pertinentes. Ces propriétés contribueront à de meilleures interactions en utilisation, particulièrement pour la variété d'interfaces ubiquitaires.

\end{translatedabstract}

\translatedkeywords{french}{\mykeywordsFR}

\maketitle

\docinfo{}

\selectlanguage{french}

\def\HEADCOLOR{ACMRed!95}
\def\headcolor{ACMRed!25}
\def\HEADERSPACE{5mm}

\newcolumntype{P}[1]{>{\raggedright\arraybackslash}p{#1}}

\newcommand{\PropTableA}[2]{
\begin{table*}
  \caption{#2}
  \label{#1}
  \begin{tabular}{|P{0.18\linewidth}|P{0.18\linewidth}|P{0.18\linewidth}|P{0.18\linewidth}|P{0.18\linewidth}|}
    \hline
    \textbf{Paramètre} & \textbf{Valeurs} & \textbf{Prisme} & \textbf{Boucle} & \textbf{Apparatus} \\
    \hline
    \parametre{} & \valeurs{} & \prisme{} & \boucle{} & \apparatus{} \\
    \hline
    \textbf{Tâche} & \textbf{Tâche de distraction} & \textbf{Faculté} & \textbf{Résultat} & \textbf{Théorie} \\
    \hline
    \tache{} & \distraction{} & \faculte{} & \resultat{} & \theorie{} \\
    \hline
  \end{tabular}
\end{table*}
}

\newcommand{\PropTableB}[3]{
{\small\sffamily
\begin{table}[#3] \caption{#2}
  \label{#1}
  \begin{tabular}{|P{0.45\linewidth}|P{0.45\linewidth}|}
    \hline
    \rowcolor{\HEADCOLOR} \multicolumn{2}{|c|}{\begin{minipage}{0.9\linewidth}\centering\vspace{\HEADERSPACE}\textcolor{white}{\normalsize\textbf{\nom{}}}\vspace{\HEADERSPACE}\end{minipage}} \\
    \hline
    \hline
    \rowcolor{\headcolor} \textbf{Prisme} & \textbf{Portée du prototype} \\
    \prisme{} & \portee{} \\
    \hline
    \rowcolor{\headcolor} \textbf{Paramètre} & \textbf{Valeurs} \\
    \parametre{} & \valeurs{} \\
    \hline
    \hline
    \rowcolor{\headcolor} \multicolumn{2}{|l|}{\textbf{Apparatus}} \\
    \multicolumn{2}{|p{0.9\linewidth}|}{\apparatus{}} \\
    \hline
    \rowcolor{\headcolor} \multicolumn{2}{|l|}{\textbf{Personnes participantes}} \\
    \multicolumn{2}{|p{0.9\linewidth}|}{\participants{}} \\
    \hline
    \rowcolor{\headcolor} \multicolumn{2}{|l|}{\textbf{Tâches}} \\
    \multicolumn{2}{|p{0.9\linewidth}|}{\taches{}} \\
    \hline
    \hline
    \rowcolor{\headcolor} \multicolumn{2}{|l|}{\textbf{Propriété}} \\
    \multicolumn{2}{|p{0.9\linewidth}|}{\resultat{}} \\
\hline
    \hline
    \rowcolor{\headcolor} \multicolumn{2}{|l|}{\textbf{Faculté humaine}} \\
    \multicolumn{2}{|p{0.9\linewidth}|}{\faculte{}} \\
    \hline
    \rowcolor{\headcolor} \multicolumn{2}{|l|}{\textbf{Explication}} \\
    \multicolumn{2}{|p{0.9\linewidth}|}{\explication{}} \\
    \hline
    \rowcolor{\headcolor} \multicolumn{2}{|l|}{\textbf{Réplication}} \\
    \multicolumn{2}{|p{0.9\linewidth}|}{\replication{}} \\
    \hline
  \end{tabular}
\end{table}
}
}

\newcommand{\ficheGEO}[2]{
\def\nom{Manipulation d'interacteurs tangibles}
\def\parametre{Multiplexage + Forme}
\def\valeurs{Espace + Spécialisée (ES),\newline Espace + Générique (EG),\newline Temps + Générique (TG)} \def\prisme{Entrée}
\def\boucle{---}
\def\portee{Applicative}
\def\participants{12 spécialistes (géophysique)}
\def\apparatus{Une table interactive qui affiche une carte géographique ou des plans de coupe. Trois interacteurs tangibles (diffraction)~: une règle (ES), deux palets (EG) et un palet (TG). Un boîtier à boutons.}
\def\taches{
  T1 (TBN)~: Spécifier des lignes de coupe depuis une carte géographique.\newline
  T2 (THN, TRU)~: Reconnaître une lettre à partir des plans de coupes spécifiés depuis une carte géographique.
}
\def\faculte{Habileté motrice, Dextérité, Charge cognitive limitée}
\def\resultat{Temps de spécification d'une ligne de coupe~:\newline ES $\approx$ EG $\approx$ TG si TBN\newline ES > EG $\approx$ TG si THN}
\def\explication{La forme spécialisée correspond mieux aux caractéristiques de la tâche, permettant de rester concentré sur un raisonnement (THN), alors que la dextérité l'habiletée des mains annule cette supériorité lors d'une tâche uniquement manipulatoire (TBN).
}
\def\replication{Étend \cite[Ch. 6.1]{fitzmaurice1996phdthesis}}
\PropTableB{tab:propGEO}{#2}{#1}
}

\newcommand{\ficheARCHEO}[2]{
\def\nom{Activation de manipulations 3D bi-manuelles}
\def\parametre{Activation}
\def\valeurs{Pédalier, Boutons}
\def\prisme{Entrée}
\def\boucle{---}
\def\portee{Applicative}
\def\apparatus{Un écran affiche deux objets 3D. Deux interacteurs manuels à six degrés de liberté, activés par un pédalier au sol ou des boutons sur les interacteurs (diffraction).}
\def\participants{12 spécialistes (archéologie) + 14 collègues experts (CAO ou jeux vidéos 3D)}
\def\taches{Assembler deux formes 3D. La difficulté s'accroît, partant de forme géométriques très simples, puis un peu moins simples, et allant jusqu'à des fragments archéologiques (TDC).}
\def\faculte{Habileté motrice, Amplitude de mouvement}
\def\resultat{Richesse des gestes~: Pédalier > Boutons}
\def\explication{Devoir maintenir les boutons enfoncés restreint les capacités rotatives à celle du poignet, entrainant de petits gestes le long des axes x, y, z. Le pédalier permet d'exploiter toute liberté de rotation entre les doigts et de mieux conserver le geste d'assemblage.}
\def\replication{---}
\PropTableB{tab:propARCHEO}{#2}{#1}
}

\newcommand{\ficheENERGY}[2]{
\def\nom{Perception d'un changement de forme}
\def\parametre{Profil de vitesse}
\def\valeurs{Constant, Exponentiel, Logarithmique}
\def\prisme{Sortie}
\def\boucle{Perception périphérique}
\def\portee{Applicative}
\def\apparatus{En vision périphérique~: un anneau physique de diamètre extensible selon trois profils de vitesse (diffraction). En vision centrale~: un écran d'ordinateur. À la main~: une souris.}
\def\participants{30 collègues}
\def\taches{
  T1 (TBN)~: Détection des mouvements de l'anneau. Distraction~: Focalisation de l'attention centrale sur un compte à rebours en vision centrale.\newline
  T2 (THN)~: Perception des mouvements de l'anneau. Distraction~: Focalisation de l'attention centrale sur la restitution d'une séquence de symboles en vision centrale.
}
\def\faculte{Vision périphérique, Attention périphérique}
\def\resultat{
  Détection (TBN) : Exp. = Log. $\gtrapprox$ Const.\newline
  Perception (THN) : Exp. > Log. > Const.\newline
  Calme : Exp. = Const. > Log.
}
\def\explication{Perception de l'environnement dans le flux optique lors de la locomomotion (\cite{gibson1950perception} cité dans \cite{traschutz2012speed}).}
\def\replication{Confirme~\cite{traschutz2012speed}}
\PropTableB{tab:propENERGY}{#2}{#1}
}

\newcommand{\ficheSURGERY}[2]{
\def\nom{Physicalité des cibles pour le pointage}
\def\parametre{Physicalité}
\def\valeurs{Virtuel, Mixte, Physique}
\def\prisme{Sortie}
\def\boucle{---}
\def\portee{Applicative}
\def\apparatus{Les mouvements de tête contrôlent la position d'un curseur affiché en réalité augmentée. Le curseur est affiché devant trois types de menus~: virtuel, mixte et physique (diffraction).}
\def\participants{15 novices (réalité mixte) avec bases en mathématiques suffisantes}
\def\taches{Sélection dans un menu (TBN). Distraction (THN)~: Deux raisonnements mathématiques, parmis addition, comparaison et mémorisation de nombres. Les deux tâches sont en alternance (TDA, TRU).}
\def\faculte{Perception visuelle}
\def\resultat{Temps de sélection~: Physique > Mixte $\approx$ Virtuel\newline Préférence~: Physique > Mixte > Virtuel}
\def\explication{La meilleure performance de lecture des informations physiques provient d'un meilleur réalisme visuel des représentations physiques (\cad{} résolution, indices stéréoscopiques, indices d'accomodation, ombres et lumières, et texture) \cite{jansen2013evaluation}.}
\def\replication{Étend \cite{bailly2019head}}
\PropTableB{tab:propSURGERY}{#2}{#1}
}

\newcommand{\fichePUBLICWORKS}[2]{
\def\nom{Indices visuels pour le positionnement 3D}
\def\parametre{Indices visuels sur scène 3D}
\def\valeurs{Ombre ancrée (O), Grille (G), Ombre ancrée + Grille (OG), Aucun (A)}
\def\prisme{Sortie}
\def\boucle{---}
\def\portee{Applicative}
\def\apparatus{Des cubes sont affichés en réalité augmentée devant une salle vide. Pour aider au référencement 3D, quatres types d'indices sont affichés sur la scène~: des ombres ancées, une grille, les deux, ou aucun (diffraction).}
\def\participants{20 novices (Réalité Augmentée)}
\def\taches{
  T1 (TRU)~: marquer la position de cubes aériens et de cubes enfouis en marchant puis déposant une croix physique au sol.\newline
  T2/T3 (TBN)~: Classer six cubes 3D par altitude et distance égocentrique.
}
\def\faculte{Perception visuo-spatiale}
\def\resultat{
  Précision du marquage (TRU)~: O $\gtrapprox$ OG > G $\gtrapprox$ A\newline
  Justesse du classement (TBN)~: OG $\gtrapprox$ O > G > A
}
\def\explication{Les ombres ancrées aident à projeter la tâche dans le plan. Lors du classement, la grille facilite la comparaison entre les ombres et les ancrages. Lors du marquage, un explication possible est que les repères mentaux pris avec une grille deviennent inadéquats au moment de se déplacer sans celle-ci, alors qu'en l'absence de grille les repères mentaux sont pris relativement à l'environnement.}
\def\replication{Complète~\cite{rosales2019distance}}
\PropTableB{tab:propPUBLICWORKS}{#2}{#1}
}

\newif\ifDETAILS
\DETAILSfalse

\ifDRAFT
\todo{}
\color{red}
\begin{enumerate}
\item étude utilisatrice => étude d'utilisation
\end{enumerate}
\color{black}
\fi

\section{Introduction}

La recherche en Interaction Humain-Machine (IHM) produit de nombreux résultats empiriques, issus aussi bien de l'évaluation d'une technologie, d'une conception ou d'une application, que de la vérification d'une hypothèse ou d'une prédiction. Les mesures effectuées portent, le plus souvent, sur l'expérience d'utilisation, la performance d'exécution et éventuellement un taux d'erreur. En revanche, ces résultats empiriques sont rarement répliqués, conduisant à une surgénéralisation de résultats isolés \cite{greenberg1992weak}. Pour se donner une idée des proportions, parmi les 546 articles publiés par la conférence CHI pour l'année 2016, 240 articles (44,0\%) concernaient des études empiriques sur l'utilisation de systèmes interactifs \cite{wobbrock2016contributions}. Par contre, de 891 articles publiés par les conférences CHI et les revues ToCHI, HCI et IJHCS de 2008 à 2010, seulement 28 articles (3,1\%) s'intéressaient à la réplication de résultats antérieurs \cite{hornbaek2014once}. Or, par voie de conséquence, peu de réplication se traduit aussi par peu de méta-analyses pouvant faire la synthèse de résultats répliqués (toujours en 2016, la conférence CHI a publié seulement 10 articles de synthèse (1,8\%), méta-analyses et revues de la littérature comprises \cite{wobbrock2016contributions}). Une explication à ce constat est que les études de réplication souffrent d'être publiées pour être peu estimées, peu réussies ou peu controversées \cite{greenberg2008usability,hornbaek2014once}. Depuis ces constatations, il semble qu'à ce jour aucune proposition forte ne soit venu contrecarrer la situation.
Pourtant, répliquer des observations contribue à préciser les conditions dans lesquelles des résultats connus sont valides et à éliminer ceux ne pouvant être répliqués dans d'autres conditions que celles de l'expérimentation originale \cite{hornbaek2014once}. En effet, des observations isolées qui apparaissent dans des conditions particulières \cite{greenberg1992weak}, ne survivent pas forcément aux prototypes, aux technologies et aux tâches impliquées lors de l'étude où elles sont apparues. Du moins, la réplication permet de s'en assurer, en bornant les conditions dans lesquelles les résultats s'appliquent ou s'estompent, augmentant ainsi la confiance dans leur validité \cite{greenberg1992weak,hornbaek2014once}.
Ainsi, répliquer les observations puis les pondérer par des méta-analyses, contribuerait donc directement à renforcer les connaissances en IHM qui reposent sur des résultats empiriques.
La question abordée par cet article est donc la suivante~: comment impulser un effort communautaire permettant de générer et répliquer des observations~?

Cet article propose de répondre à cette question dans le cas d'observations du phénomène de l'interaction, en se focalisant sur les boucles d'interaction impliquées dans les systèmes interactifs. L'approche proposée est de greffer, systématiquement, aux étapes de traitement de cas applicatifs, des études expérimentales additionnelles permettant d'observer une propriété interactionnelle, tout en optimisant le prototype applicatif. Cet article propose donc de ne plus produire des résultats empiriques uniquement depuis des tâches génériques de bas niveau, mais aussi à partir de tâches de plus haut niveau, telles que les tâches de personnes utilisatrices finales variées. Pour ce faire, cet article définit les propriétés interactionnelles et en fournit un formalisme de description, un outil d'observation (les prismes interactionnels) et une méthode d'observation calibrée (les études de propriétés).

Les observations ainsi générées peuvent aussi bien servir à informer immédiatement les besoins de conception en ingénierie, qu'à participer à l'établissement de lois, à la définition de règles ou à la construction de théories. Notamment, observer l'interaction, plutôt que les interfaces, est particulièrement fondé dans le domaine des interfaces ubiquitaires, qui relève d'une grande diversité de technologies et de contextes d'usages possibles, et où la forme des interfaces est tout sauf universelle (contrairement aux interfaces graphiques utilisant un écran, un clavier et un dispositif de pointage tel qu'une souris) et évolue constamment en s'adaptant aux cas d'applications et au renouvellement des technologies.
Aussi, se greffer au traitement de cas applicatifs relève d'un double intérêt. Premièrement, cette greffe est un moteur de diversification des conditions d'observations (\pex{} prototypes, technologies, tâches et profils de personnes utilisatrices), ce qui est bénéfique lors de la réplication de résultats. Cette approche est donc fondée scientifiquement. Deuxièmement, cette greffe est moteur d'impulsion pour la communauté, car une fois que les tâches des personnes utilisatrices finales ont été analysées et que des prototypes ont été développés pour répondre aux besoins de traitement du cas applicatif, adjoindre une étude de propriété aux études d'utilisabilité usuelles demandera un effort et un budget additionnels négligeables. De plus, ces efforts additionnels, pour greffer une étude de propriété au plan de traitement d'un cas applicatif, sont récompensés car les résultats obtenus optimisent le prototype applicatif, permettent une meilleure justification des choix de conception et une meilleure compréhension des raisons du bon fonctionnement du prototype, tout en facilitant la publication par l'ajout d'une étude plus informative. Cette approche est donc aussi fondée économiquement. Ainsi, cette greffe permet aux communautés en IHM de collecter la réplication d'un résultat connu ou d'un nouveau résultat qui sera à répliquer, contribuant ainsi à l'effort scientifique de la communauté sans avoir dû y dédier des efforts considérables de développement et de financement. L'approche proposée, qui permet d'enrichir de simples cas applicatifs, est donc à même de pouvoir répondre au besoin d'impulsion à l'échelle communautaire par une double reconnaissance technologique et scientifique, ainsi qu'au besoin de résultats répliqués.

L'article est structuré de la manière qui suit. Tout d'abord, l'article commence par établir le périmètre du contexte de recherche dans lequel il intervient. Ensuite, l'article fait plusieurs apports, sur la caractérisation des études expérimentales, sur la caractérisation des conditions expérimentales, mais aussi par l'imagination d'un outil de révélation des boucles d'interaction, ainsi que par la définition de propriétés interactionnelles, accompagnées d'un formalisme de description. L'article propose alors son approche visant à observer l'interaction en optimisant des prototypes applicatifs, par ajout d'études de propriétés interactionnelles à la chaîne de traitement des cas applicatifs, qui est ensuite illustrée par cinq cas. L'article conclue par une discussion sur le potentiel et la viabilité de l'approche en matière de réplication, et par des directions pour les futures recherches.

\section{Contexte de la recherche}\label{sec-related-work}

En IHM, les études impliquant des personnes utilisatrices peuvent aussi bien intervenir dans un cadre expérimental contrôlé (\cad{} étude expérimentale), que dans un cadre observationnel au cours de l'utilisation, de manière écologique (\cad{} étude d'utilisation). À cet effet, différents buts peuvent être visés, selon différentes approches. Néanmoins, les études expérimentales sont souvent isolées les unes des autres de par les prototypes, les technologies et les tâches impliquées, et les études de réplication restent trop peu nombreuses \cite{cockburn2020threats,greenberg1992weak,greenberg2008usability,hornbaek2014once,hornbaek2015wrong,molich2018usability}, conduisant à une surgénéralisation de résultats empiriques isolés \cite{greenberg1992weak}. Or, les résultats de ces études expérimentales concernent le plus souvent l'utilisabilité de prototypes, la performance d'une technologie ou la réussite de tâches, délaissant l'exploration de l'interaction, qui est pourtant à l'\oe{}uvre dans chacun de ces cas. En outre, les conditions expérimentales restent difficiles à exprimer et à comparer, limitant de fait la possible intégration de résultats éparses. Dans ce contexte de recherche, cette section revient sur les types d'expérimentations par utilisation, la réplication de résultats empiriques, les découvertes fortuites dans les sciences, l'étude de l'interaction et la description des tâches interactives.

\subsection{Les expérimentations par utilisation}

Une approche classique aux sciences empiriques est la \textit{recherche expérimentale}, reposant sur des expérimentations testant des relations causales. Cette approche s'inscrit dans la \textit{recherche quantitative}, examinant les relations de \textit{causalité} entre des \textit{variables} prédites par des \textit{hypothèses} \cite{gergle2014experimental} (les prédictions émanant des connaissances théoriques)~: les expérimentations produisent alors les preuves qui valident ou invalident les hypothèses. En IHM, de telles expérimentations impliquent des personnes utilisatrices et des interfaces humain-machine. Or, la variabilité du comportement humain complique les mesures expérimentales \cite{gergle2014experimental}, de même que la variété des interfaces humain-machine et des dispositifs d'interaction interrogent les conditions d'apparition, nécessitant de répéter les mesures au travers de multiples conditions d'interaction, ainsi que de reproduire les résultats par des équipes de recherche indépendantes \cite{gergle2014experimental,greenberg1992weak,hornbaek2014once}. S'inscrivant aussi bien dans les sciences sociales (telles que l'ethnographie \cite{mackay1997triangulation,mackay1999strips}), l'IHM recourt également à des méthodes de la \textit{recherche qualitative}. Cette approche vise la compréhension subjective de phénomènes difficiles à quantifier, plutôt que la mesure expérimentale obtenue en conditions contrôlées \cite{adams2008qualitative}. La combinaison des méthodes quantitatives et qualitatives se retrouve dans les approches de la \textit{recherche mixte} \cite{johnson2004mixed}. Le croisement de ces méthodes en IHM apparaît, par exemple, lors de l'interprétation de commentaires des personnes utilisatrices issus de verbalisations ou de questions ouvertes, alors que des mesures d'exécution renseignent déjà sur la performance d'utilisation d'un dispositif interactif.

L'IHM peut recourir distinctement à des approches aussi bien quantitatives que qualitatives, selon le problème soulevé. Conduire des expérimentations par utilisation en conditions contrôlées (\cad{} approche quantitative) répond essentiellement à trois besoins \cite{hornbaek2013whys}~: valider une technologie (B1), répondre à une question de recherche (B2) et rationaliser une conception (B3), c'est-à-dire respectivement des problèmes constructifs, conceptuels et empiriques \cite{oulasvirta2016solving}. Cependant, ces mêmes problèmes peuvent impliquer des phénomènes ne pouvant être observés en conditions contrôlées ou ne nécessitant pas la rigueur de leurs protocoles \cite{hornbaek2013whys}, tels que~: l'adoption d'une technologie (B4), préférant des études de terrain ou des entretiens qualitatifs, l'observation d'un phénomène trop difficile à caractériser (B5) ou encore l'information de choix de conception (B6), préférant de brèves études d'utilisabilité ou de longues études de cas par itérations. Le présent article focalise les trois premiers besoins susmentionnés (\cad{} B1, B2 et B3) et leur combinaison possible au sein d'une même recherche, résolvant alors un \textit{problème mixte} \cite{oulasvirta2016solving}. Toutes les six paires sont effet possibles entre des problèmes constructifs, conceptuels et empiriques \cite{oulasvirta2016solving}. En particulier, au sein d'une même recherche, des expérimentations par utilisation testant une technologie ou une conception peuvent aussi servir à établir des théories dépassant ce cadre (\cad{} combinaison d'un problème constructif ou empirique avec un problème conceptuel). L'approche défendue dans cet article entre dans cette voie. Néanmoins, la variabilité du comportement humain, de même que la variété des interfaces humain-machine et l'évolution constante des dispositifs d'interaction, nécessitera de répéter les mesures avant de permettre toute généralisation. Ainsi, la section suivante revient sur les pratiques de réplication des résultats en IHM.

\subsection{La réplication de résultats empiriques}

Les sciences empiriques reposent sur des observations expérimentales. Les conclusions ainsi obtenues doivent pouvoir être reproduites de manière indépendante~: un même protocole et des conditions expérimentales identiques doivent pouvoir retrouver les mêmes observations et aboutir aux mêmes conclusions. Ainsi, la \textit{réplication stricte} permet de vérifier la validité de recherches déjà publiées \cite{hornbaek2014once}. De la sorte, ces dernières années, plusieurs sciences empiriques traversent une crise de la reproductibilité des résultats, notamment lorsque la nécessité de publier oriente les recherches ou interfère avec l'obtention de résultats \cite{cockburn2020threats}.

Cependant, le but de la réplication est parfois autre que de chercher à vérifier des résultats, comme de chercher à les étendre ou à mieux borner leurs conditions d'apparition. Par exemple, la \textit{réplication partielle} reprend seulement une partie des facteurs étudiés et en altère d'autres, la \textit{réplication conceptuelle} suit une procédure complètement différente de l'étude originale, et la \textit{réplication systématique} combine une réplication stricte avec des changements de variables \cite{hornbaek2014once}.

Pouvoir générer et accumuler des observations variées, répliquées au travers de multiples technologies, applicatifs ou tâches, peut s'avérer structurant pour la science de l'IHM. En effet, de telles réplications permettraient de faire émerger des résultats s'émancipant des technologies, des applicatifs et des tâches. Aboutir à de tels résultats transversaux semble particulièrement pertinent alors que les interfaces humain-machine sont devenues ubiquitaires, multimodales, et de multiples sortes \citetext{\citealp[Ch.~20]{dix2003hci},\citealp{quigley2010uui},\citealp{oviatt2017handbook},\citealp[Ch.~7]{sharp2019beyond}}, que les technologies utilisées se renouvellent sans cesse, et que ces mêmes technologies peuvent se voir utilisées pour de multiples applications et pour diverses tâches, et que de mêmes applications ou tâches peuvent se retrouver au travers de technologies variées. Ainsi, cet article s'intéresse à la réplication dans ce second but, avec pour ambition de pouvoir faire émerger des résultats récurrents et de préciser les conditions dans lesquelles ils s'appliquent.

En revanche, généraliser des résultats à partir d'observations multiples et variées nécessitera de pouvoir considérer et pondérer les différentes conditions d'observation, ainsi que les diverses conclusions. Une réponse à cette nécessité réside dans les méta-analyses.

\subsubsection{Intégration et généralisation des résultats}

La logique de réplication atteint sa finalité lorsque des analyses ultérieures peuvent pondérer, synthétiser, voire généraliser les résultats empiriques accumulés. À cette fin, les méta-analyses relèvent d'une approche quantitative d'intégration de résultats empiriques, et doivent donc être différenciées d'autres approches de synthèse ou de collecte de recherches, telles que les revues systématiques de la littérature ou les sondages de recherches. De même, d'autres méthodes de synthèses relèvent d'approches qualitatives \cite{hansen2022guide}, telle que la \textit{méthode de sondage d'études de cas}, qui produit des résultats qualitatifs à partir de l'intégration de recherches qualitatives \cite{jurisch2013case}.

Les méthodes de méta-analyses sont utilisées depuis au moins un siècle \cite{sutton2008recent}, ont été documentées (\pex{} \cite{cooper2009handbook,haidich2010meta,hansen2022guide,sutton2008recent}) et sont, par exemple, très fréquemment utilisées dans le domaine médical \cite{sutton2008recent,thacker1988integration}, pouvant même parfois devenir redondantes, contradictoires, ou plus nombreuses que les recherches originales \cite{chapelle2021epidemic,sigurdson2020redundant}. Loin de cette profusion, le domaine de l'IHM a néanmoins déjà commencé à produire des méta-analyses. Par exemple, en répondant à des questions sur des technologies (B1), comme d'évaluer le meilleur rôle à donner aux écrans vibrotactiles \cite{chai2022vibrotactile}, sur des questions de recherche (B2), comme d'établir le lien entre la personnalité humaine et l'acceptation d'un robot \cite{esterwood2021personality}, et aussi sur des questions de conception (B3), comme d'établir l'influence de l'esthétique sur la performance \cite{thielsch2019aesthetics}. Cependant, ces initiatives restent clairsemées. Aucune stratégie communautaire de réplication ne permet encore de générer une masse suffisante de réplications qui justifierait l'essor de telles pratiques en IHM. Ainsi, une originalité de cet article est de proposer une méthode pouvant impulser un effort à l'échelle communautaire.

La généralisation permet d'étendre des résultats à des contextes plus larges que les conditions spécifiques de technologies, d'interfaces humain-machine et de tâches, dans lesquelles ils ont été constatés \cite{sutcliffe2000reuse}. Que ce soit lors d'une expérimentations par utilisation, ou encore plus lors d'une méta-analyse, la généralisation renforce donc la valeur du résultat. Pour pouvoir généraliser un résultat, il est nécessaire de pouvoir en donner une explication. Cette explication peut être antérieure à l'expérimentation, sous la forme d'une hypothèse préalable, le plus souvent appuyée par une théorie préexistante. Mais l'explication peut parfois aussi être postérieure, en l'absence de théorie préexistante. Néanmoins, dans l'un ou l'autre cas, la généralisation est possible uniquement si une théorie sous-jacente est rattachée à l'explication \cite{greenberg1992weak}. De nombreuses théories sont liées à la compréhension des interactions humaines avec les technologies \cite{mbl2021generative} et peuvent permettre d'émettre des hypothèses anticipant les phénomènes survenant lors d'une expérimentations par utilisation. Somme toute, il est aussi possible de travailler sans théorie ou hypothèse préalable. Cette approche, qui est à différencier de la modification d'hypothèses pour obtenir la signification statistique \cite{cockburn2020threats}, s'inscrit dans la sérendipité, comme décrit dans la section suivante.

\subsection{Les découvertes fortuites dans les sciences}

\def\bigbangstoryfootnotetext{En 1964, Arno Penzias et Robert Wilson (astronomes aux laboratoires Bell) devaient réemployer, pour l'observation spatiale, une antenne radio précédemment utilisée pour la communication avec des satellites, en la transformant en radiotélescope à faible bruit. Cependant, des bruits résiduels continuaient d'être détectés après avoir effacé les signaux terrestres puis nettoyé l'antenne \citetext{\citealp{durrer2015bigbang}, \citealp[Ch.~19]{roberts1989serendipity}}. À la même période, le groupe de recherche mené par Robert Dicke, sur la gravitation et la relativité (à l'université de Princeton), travaillait sur la théorie du Big Bang, en cherchant un bruit résiduel suspecté d'exister \citetext{\citealp[Ch.~19]{roberts1989serendipity}, \citealp{wilkinson2000bigbang}}. Après près d'un an de questionnements, Penzias fut informé des investigations de Dicke lors d'un appel téléphonique avec Bernard F. Burke (radioastronome au MIT). Alors, Penzias rentra en contact avec Dicke, qui fit le déplacement aux Bell Labs avec Peter Roll et David Wilkinson pour examiner les données et les détails expérimentaux \citetext{\citealp{durrer2015bigbang}, \citealp{wilkinson2000bigbang}}. Treize ans plus tard, Penzias et Wilson reçurent le prix Nobel, en 1978, pour la découverte du bruit de fond cosmique (CMBR). Dans l'équipe de Dicke, James Peebles a reçu un prix Nobel en 2019.}

La sérendipité, qui fut décrite pour la première fois par Horace Walpole en 1754 \cite{yaqub2018taxonomy}, survient lorsque les ``\textit{scientifiques font des découvertes inattendues mais profitables}''\cite{yaqub2018taxonomy} (\cad{} le caractère parfois imprévisible de la recherche scientifique). La science recèle de nombreux cas de sérendipité \cite{chevrier2018caalors,roberts1989serendipity}, tels que le cas emblématique de Arno Penzias et Robert Wilson qui ont découvert la première preuve de la théorie du Big Bang alors qu'ils étaient chargés de réemployer une ancienne antenne radio\footnote{\bigbangstoryfootnotetext{}} \citetext{\citealp[Ch.~12]{chevrier2018caalors}, \citealp{durrer2015bigbang}, \citealp[Ch.~19]{roberts1989serendipity}, \citealp{wilkinson2000bigbang}}. Malgré le fait que cette théorie leur était totalement inconnue, leur curiosité et leur ténacité ont été clefs pour établir le lien entre leur observation et la théorie \ifDRAFT\rouge{\cite[Ch.~11]{chevrier2018caalors}}\else\cite[Ch.~11]{chevrier2018caalors}\fi. Ce cas nous enseigne que la voie directe -- entre une prédiction théorique et des éléments factuels -- est parfois trop longue, lorsque des cas applicatifs font apparaître des observations inattendues et inexpliquées de manière inopinée.

\def\serendipitykindsfootnotetext{Les quatre cas de sérendipité sont~: (1)~\textit{Walpolienne}~: ``Une recherche ciblée résout un problème inattendu''; (2)~\textit{Mertonienne}~: ``Une recherche ciblée résout le problème en question par une voie inattendue''; (3)~\textit{Bushienne}~: ``Une recherche non-ciblée résout un problème immédiat''; et (4)~\textit{Stephanienne}~: ``Une recherche non-ciblée résout un problème ultérieur'' \cite{yaqub2018taxonomy}.}

\def\serendipitymechanismsfootnotetext{Les quatre mécanismes de sérendipité sont~: (1)~\textit{Dirigée par la théorie}~: ``Le développement de la théorie rend la sérendipité évidente à tout observateur''; (2)~\textit{Dirigée par l'observateur}~: ``La sérendipité est apparente seulement à certains par leurs outils, techniques et aptitudes''; (3)~\textit{Émanant d'une erreur}~: ``La sérendipité peut émerger à la suite d'écarts méthodologiques, d'erreurs et de maladresses''; et (4)~\textit{Émergeant par réseau}~: ``La sérendipité peut impliquer un réseau d'acteurs'' \cite{yaqub2018taxonomy}.}

La sérendipité est tout sauf le fruit du hasard \cite{yaqub2018taxonomy}~: elle arrive aux scientifiques préparés après un travail acharné \cite{abelson1963serendipity}. Ce phénomène est même étudié scientifiquement~: par exemple, une taxonomie définit quatre genres\footnote{\serendipitykindsfootnotetext} et quatre mécanismes\footnote{\serendipitymechanismsfootnotetext} de sérendipité \cite{yaqub2018taxonomy}, concluant que la pression accrue et le cadrage intensif de la recherche et de l'innovation réduit les opportunités d'apparition de la sérendipité \citetext{\citealp[Ch.~16]{chevrier2018caalors}, \citealp{yaqub2018taxonomy}}. Les cas de sérendipité doivent donc pouvoir encourager les recherches en IHM à être pensées différemment, en dépassant l'attente de théories établies, pour conduire des expérimentations par utilisation visant à générer ou répliquer des résultats conceptuels (B2). Ainsi, l'approche proposée dans cet article repose en partie sur la sérendipité.

\subsection{L'interaction comme objet d'étude}

Le terme ``interaction'' a de plus en plus reçu l'attention des recherches en IHM au fil du temps \cite{hornbaek2019mean}, par un glissement de la perspective prise sur le champ d'étude, pour passer de la création ``d'interfaces humain-machine'' à la création ``d'interactions''  \cite{mbl2004designing}. Or, la perspective prise pour un objet d'étude est conséquente dans le sens où elle conditionne son observation et son interprétation, mais aussi les moyens d'évaluation et de conception \cite{hornbaek2017interaction}. Une récente définition de ce terme indique que  ``l'interaction`` survient lorsque ``\textit{deux entités déterminent mutuellement leurs comportements respectifs au cours du temps}'' et que, en l'occurrence, ces entités sont un système interactif et une personne utilisatrice \cite{hornbaek2017interaction}. Une des façons de voir cette interaction est celui d'un ``\textit{dialogue composé d'échanges cycliques, passant par des canaux d'entrées et de sorties (du point de vue du système interactif) ou des canaux d'actions et de perceptions (du point de vue de la personne utilisatrice)}'' \cite{hornbaek2017interaction}.

Cette perspective prise sur l'interaction, vue comme un dialogue, permet alors une caractérisation particulière. Ainsi, selon la théorie de l'action, ces échanges cycliques consistent, pour la personne utilisatrice, en sept étapes \cite{norman2002design}. De plus, plusieurs distances apparaissent de part et d'autre des cycles, au sein du gouffre de l'exécution et du gouffre de l'évaluation, et sont à réduire pour obtenir une meilleure interaction \cite{norman2002design}.
Enfin, comme ces échanges surviennent de manière dynamique, au cours du temps, l'interaction peut se voir comme un phénomène, se traduisant par une alternance d'échanges au cours de cycles successifs. Dans la littérature, les différents cycles possibles avec une interface numérique sont le plus souvent retrouvés sous l'appellation de ``boucles'' voire de ``boucles de rétroaction''. Étudier les boucles d'interaction est alors un moyen d'étudier l'interaction.

\subsubsection{Les boucles d'interaction}

Plusieurs boucles d'interaction ont déjà été caractérisées dans des théories ou des modèles d'interaction. Par exemple, le modèle de l'interaction instrumentale \cite{mbl2000instrumental} permet d'observer trois boucles après une \textit{action} sur la partie physique d'un instrument~:
\begin{enumerate}
\item Une première boucle de \textit{réaction} de la partie physique de l'instrument. Par exemple, la souris a changé physiquement de position par action de la personne utilisatrice.
\item Une deuxième boucle de \textit{réaction} de la partie logique de l'instrument. Par exemple, la glissière de l'ascenseur a changé de position.
\item Une troisième boucle de \textit{réponse} de l'objet cible de l'instrument. Par exemple, la vue du document est réajustée.
\end{enumerate}

La modélisation de l'interaction avec des interfaces tangibles a quant à elle précisé jusqu'à trois types de boucles de rétroactions possibles \cite{ishii2008beyond}~:
\begin{enumerate}
\item Une première boucle de \textit{rétroaction physique immédiate}. Par exemple, un objet a changé de position par action de la personne utilisatrice.
\item Une deuxième boucle de \textit{rétroaction numérique post-calcul}. Par exemple, une représentation graphique associée à l'objet est modifiée en conséquence.
\item Une troisième boucle de \textit{rétroaction physique par actuation}. Par exemple, l'ordinateur maintient une représentation physique en cohérence avec l'évolution de l'état d'une donnée.
\end{enumerate}

Les boucles d'interaction se caractérisent donc par une entité émettrice, une entité perceptrice et le type de l'information véhiculée. Aussi, le type de boucle et le nombre de boucles dépend de l'interface numérique et du modèle d'interface auquel elle répond. Cependant, la littérature semble ne pas recéler de méthode spécifique à l'observation de ces boucles, ou même de présentation systématique ou de projection des résultats en fonction de la boucle impliquée. Pourtant, étudier ces boucles seraient un moyen de procéder à l'étude de l'interaction, plutôt qu'à l'étude des interfaces humain-machine.

Selon la perspective retenue dans la présente section, les études expérimentales, conduites avec des systèmes interactifs, mettent donc à l'épreuve des échanges cycliques, qui empruntent ou activent des boucles d'interaction spécifiques, dépendantes à chaque interface numérique et à chaque paradigme. Une originalité du présent article repose sur la proposition de spécifier les résultats expérimentaux dans les termes de l'interaction, et plus précisément de l'interaction considérée comme un dialogue, et d'adapter les méthodes d'observation à cet effet. L'objet d'étude devient alors le phénomène d'interaction, plutôt que les interfaces humain-machine. Ce glissement a également pour but de modifier l'observation et l'interprétation de l'interaction, tout comme les moyens de concevoir et d'évaluer l'interaction.

\subsection{La description de tâches interactives}

Les méthodes, les modèles et les notations pour l'analyse des tâches interactives \cite{adams2012choosing,bowen2021task,diaper2004understanding} permettent de décrire les tâches d'une personne utilisatrice avec un système interactif. Les modèles de tâches peuvent servir deux grands rôles. Tout d'abord, certains de ces modèles servent un rôle prescriptif en spécifiant uniquement les tâches interactives, sans se préoccuper du système interactif, et en permettant ensuite d'appuyer les phases de conception et de développement \cite{bowen2021task}. D'autres modèles jouent un rôle descriptif, en décrivant les tâches permises par un système interactif déjà existant, et en permettant de soutenir des évaluations \cite{bowen2021task}. Par exemple, de premières évaluations, telles que des études d'utilisabilité, peuvent être conduites avec des prototypes dès que la conception du système interactif, ou même d'une sous-partie de ce système, est connue avec suffisamment de détails \cite{veer2004dutch}. Parmi tous les modèles disponibles, une constante reste le soin apporté à la finesse de description des hiérarchies des tâches et de leurs relations temporelles \cite{bowen2021task}. Cependant, ces modèles qui sont suffisamment élaborés et opérationnels pour assurer la conception précise et complète d'une totalité de tâches, ainsi qu'une ébauche et une élaboration de systèmes d'envergure, deviennent surdimensionnés pour cerner des cas applicatifs dont le périmètre se restreint à démontrer l'utilité d'une technologie, comme souvent rencontré lors de recherches explorant, par exemple, le vaste espace de conception offert par les interfaces ubiquitaires. D'une manière plus large, le manque ressenti concerne la description de tâches dans le cadre de protocoles expérimentaux, de sorte à simplement décrire les conditions expérimentales d'obtention d'un résultat empirique, plutôt que la spécification complète ou l'analyse d'un système interactif.

En effet, lors de conditions expérimentales, le prototype évalué dépasse rarement un morceau d'interface numérique ou une sous-partie d'une application, spécifiquement développé pour le besoin de la recherche en cours. Étoffer la conception à l'échelle d'un système complet et à haut niveau de détail est alors peu pertinent, et les outils à disposition sont surdimensionnés. Apprendre à utiliser ces outils en profondeur pour en devenir suffisamment familier semble disproportionné. Par exemple, pouvoir combiner de nombreux niveaux de hiérarchie et des relations temporelles complexes dépasse le cadre de protocoles testant le plus souvent une ou deux tâches, tout au plus composées d'une sous-tâche. En outre, retrouver une pluralité de modèles possibles \cite{bowen2021task}, distribuée au sein de nombreuses recherches, rendrait plus difficile la comparaison des conditions expérimentales entre les résultats. Il existe donc un intérêt à choisir les éléments pertinents à la description des tâches impliquées dans le cadre restreint des protocoles expérimentaux, aux critères de choix de ces tâches ou sous-tâches, et dans le simple but de décrire les conditions d'obtention d'un résultat empirique.

\section{Une caractérisation des études expérimentales}\label{sec-usm}

\begin{figure}\centering
\includegraphics[trim={1.6cm 1.8cm 1.6cm 1.2cm},clip,width=8.47cm]{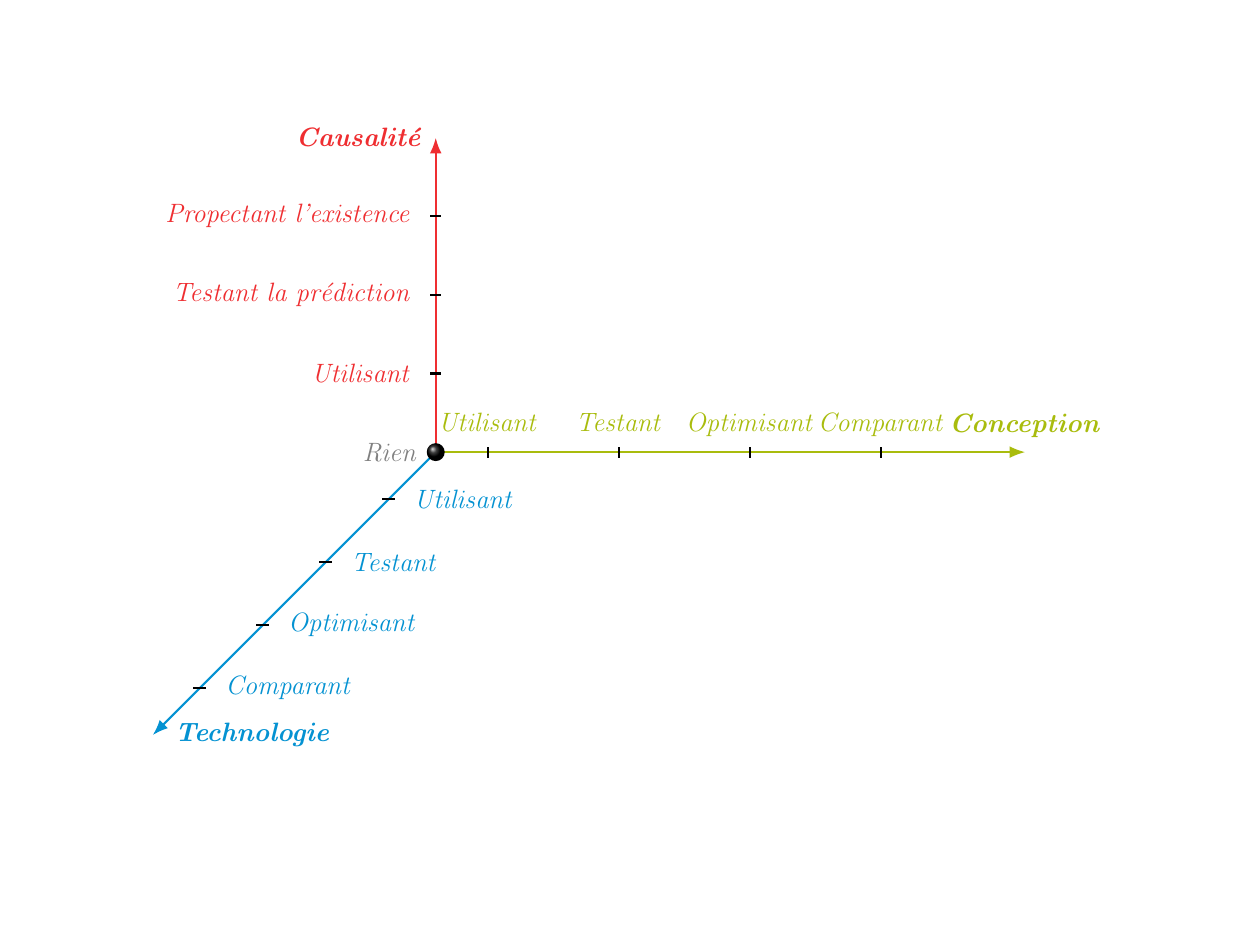} \caption{Les trois facettes du cadre conceptuel de motivation des études expérimentales (UEM). Une même étude d'utilisation peut receler plusieurs facettes.}
  \Description{The user study framework comprises three dimensions: technology, causality, and design; each is represented in 3D space by an orthogonal axis. The technology axis contains two values: `testing' and `comparing'. The causality axis contains two values: `testing prediction' and `hunting existence'. The design axis contains two values: `testing' and `comparing'. The origin value of the three axes is `None'.}
  \label{fig:studies-framework}
\end{figure}

Trois raisons motivent les recherches en IHM à produire des preuves empiriques au travers d'expérimentations par utilisation \cite[Ch.~2]{hornbaek2013whys}~: valider une technologie (B1), répondre à une question de recherche (B2) et rationaliser une conception (B3). Ces raisons pouvant se combiner, dans un problème mixte \cite{oulasvirta2016solving}, conduisent ainsi à trois facettes possibles des études expérimentales. Une caractérisation des études expérimentales, au travers de ces trois facettes, est définie par le \textit{cadre motivationnel des études expérimentales} (UEM~: \textit{User Experiment Motivation framework}) que nous proposons en \autoref{fig:studies-framework}. La suite de la présente section détaille chacune des trois facettes de cette caractérisation, en illustrant avec des exemples issus de la littérature couvrant des variétés de technologies, de tâches et d'hypothèses, provenant de revues (1 article\footnote{Les illustrations incluent 1 article de revue issu de Journal of Motor Behavior.}) et de conférences en IHM (29 articles\footnote{Les illustrations incluent 29 articles de conférences~: 8 provenant de ACM CHI, 5 de ACM UIST, 5 de ACM TEI, 2 de ACM CHI EA, 2 de ACM ITS/ISS, 2 de IFIP INTERACT, 2 de ACM AVI, 1 de ACM DIS, 1 de ACM SAP et 1 de ACM CHINZ.}), sélectionnés à partir d'études déjà connues \ifACCEPTED {de l'auteur} \else {des auteurs} \fi (12 articles) et complétées par des collections obtenues par recherches de mots-clefs\footnote{Les mots-clefs utilisés portaient des études expérimentales (\pex{} ``User Study'', ``User Experiment'', ``Evaluation''), des cas applicatifs et des études de cas (\pex{} ``Case Study'', ``Use Case'', ``Application''), et des technologies et des styles d'interaction qui ont fait l'objet d'études durant les décennies passées (\pex{} ``Augmented Reality'', ``Tabletop'', ``Multi-touch'').} dans la base bibliographique ACM Digital Library et le moteur de recherche Google Scholar, puis systématiquement analysés manuellement. Les recherches et analyses étaient interrompues pour un type d'étude dès que pouvaient satisfaire trois exemples illustratifs, suffisamment discriminants entre eux, et lorsque possible, issus de conférences et journaux différents, et de périodes variées.

\subsection{La facette technologie}

Une étude peut simplement utiliser des technologies, sans chercher à les étudier (\cad{} sans les mettre en question). Sinon, nous distinguons les études cherchant à tester, à optimiser ou à comparer des technologies.

\subsubsection{Les études testant la technologie} 

Les \textit{études expérimentales testant la technologie} évaluent si une nouvelle technologie, développée et proposée dans des recherches, est utilisable (\pex{} accomplissement des tâches, taux d'erreur et performance de la tâche) \citetext{\citealp{greenberg2008usability}, \citealp[Ch.~2]{hornbaek2013whys}}. Les tâches mesurées sont typiquement des tâches de bas niveau spécifiques aux tests technologiques (\pex{} le pointage multi-directionnel \cite{iso:9241-9:2000}), ne nécessitant pas forcément de concevoir une technique d'interaction ou une application lorsqu'il s'agit uniquement de tester un dispositif interactif. Trois illustrations de la littérature sont, par exemple~:
\begin{enumerate}[label=(E\arabic*)]
\item Évaluer la réussite de tâches de lecture d'informations de bas niveau en utilisant un histogramme physique \cite[Étude expérimentale~1]{daniel2019cairnform};
\item Tester la performance de pointage d'une bague d'entrées 2D sur toutes surfaces, en conditions assise ou debout \cite{kienzle2014lightring};
\item Tester la compensation de latence et de précision d'un écran tactile à faible latence, au travers de tâches de dessin, d'écriture et de tirage \cite{shultz2022tribotouch}.
\end{enumerate}

\subsubsection{Les études optimisant (ou explorant) les technologies} 

Les recherches en IHM peuvent aussi recourir à des \textit{études expérimentales optimisant (ou explorant) des technologies}, en faisant varier une caractéristique technologique et en évaluant les impacts induits (\pex{} sur les temps d'exécution, les taux de réussite et les préférences d'utilisation). Trois illustrations de la littérature sont, par exemple~:
\begin{enumerate}[label=(E\arabic*)]\setcounter{enumi}{3}
\item Explorer l'assignation des doigts, l'assignation des indices et les palettes de couleur pour une technique de pointage 3D \cite{delamare2022multifingerbubble};
\item Explorer l'amplitude de deux dispositifs lors de la reconnaissance d'objets pour des formes restituées de manière haptique \cite{gonzalez2021xrings};
\item Optimiser la précision d'une surface sensible au toucher pour trois conditions de signal et trois assignations des doigts \cite{oh2021fingers}.
\end{enumerate}

\subsubsection{Les études comparant les technologies} 

Le plus souvent, lorsque possible, les études de technologies sont des études comparatives qui visent à démontrer le gain obtenu par rapport aux précédentes conditions de base et aux technologies les plus compétitives. Trois illustrations de la littérature, des \textit{études expérimentales comparant des technologies}, sont les suivantes~:
\begin{enumerate}[label=(E\arabic*)]\setcounter{enumi}{6}
\item Comparer les interactions tactiles multi-points et tangibles pour une tâche représentative de la tâche des personnes utilisatrices finales \cite{al-megren2016comparing};
\item Comparer la performance des entrées souris+clavier, multi-points et tangible pour des tâches de placements 3D \cite{besancon2017mouse};
\item Comparer des boutons molette tactiles multi-points et tangibles pour des tâches de ciblage \cite{voelker2015knobology}.
\end{enumerate}

\subsection{La facette causalité}

Une étude peut simplement impliquer des causalités, sans chercher à les étudier (\cad{} sans les mettre en question). Sinon, nous distinguons les études cherchant à tester des prédictions ou à prospecter l'existence de causalités.

\subsubsection{Les études testant la causalité} 

Les \textit{études expérimentales testant la causalité} (ou \textit{expérimentations de vérification d'hypothèse}) visent à répondre à des questions de recherche \cite[Ch.~2]{hornbaek2013whys}. Les questions de recherches bien posées visent des résultats théoriques qui survivent aux prototypes, aux technologies et aux tâches impliquées, en testant des hypothèses ou, entre d'autres termes, des relations de causalité entre des variables prédites par des théories (\cad{} évaluation des prédictions de causalité). Les tâches testées sont habituellement de bas niveau (comme des tâches génériques, des tâches élémentaires, ou de simples actions), qui ne répondent pas à des contraintes émanant de personnes utilisatrices finales ou d'un applicatif particulier. Les tests de causalité peuvent parfois s'effectuer sans même recourir à un système interactif, en démontrant des phénomènes par des expérimentations non-numériques (\pex{} compter des coups de stylo grâce à des circuits de tubes à gaz \cite{fitts1954law}). Parmi les illustrations de la littérature comptent les trois suivantes~:
\begin{enumerate}[label=(E\arabic*)]\setcounter{enumi}{9}
\item Tester les prédictions de performance de menus \cite{cockburn2007predictive};
\item Tester les prédictions de préférences d'assignation des mains pour des tâches asymétriques \cite{guiard1987asymmetric};
\item Tester les prédictions de performance pour le suivi de cibles selon le multiplexage et la forme sur une interface numérique palpable \cite{fitzmaurice1997empirical}.
\end{enumerate}

\subsubsection{Les études prospectant la causalité} 

Travailler sans hypothèses préalables est tout à fait possible (\cad{} la ``chasse au phénomène'' \cite{fromkin1976laboratory}, citée dans \cite[Ch.~4]{hornbaek2013whys}), en cherchant une relation entre des variables sans avoir une théorie pouvant l'expliquer (\cad{} évaluer l'existence d'une causalité potentielle). Des \textit{études expérimentales prospectant la causalité} trouvées dans la littérature incluent les trois illustrations suivantes~:
\begin{enumerate}[label=(E\arabic*)]\setcounter{enumi}{12}
\item Observer les effets de la facilité d'utilisation et de l'esthétique sur les préférences d'utilisation \cite{deangeli2006influence};
\item Observer les effets de grilles sur la mémoire spatiale et la mémoire des contenus \cite{leifert2011grids};
\item Observer les effets de la taille d'artefacts sur l'expérience d'utilisation \cite{lopez2021scaling}.
\end{enumerate}

\subsection{La facette conception}

Une étude peut simplement impliquer une conception (\cad{} pour répondre à un besoin), sans chercher à l'étudier (\cad{} sans la mettre en question). Sinon, nous distinguons les études cherchant à tester, à optimiser ou à comparer des conceptions.

\subsubsection{Les étude testant la conception} 

Les \textit{études expérimentales testant la conception} assurent que les interfaces humain-machine se comportent comme prévu et correspondent aux besoins des personnes utilisatrices finales et aux besoins applicatifs \citetext{\citealp{greenberg2008usability}, \citealp[Ch.~2]{hornbaek2013whys}} et mesurent ``\textit{ce qui fonctionne et ne fonctionne pas dans une interface}'' \cite[Ch.~2]{hornbaek2013whys}. Les recherches peuvent recourir à ces études lors de la conception de prototypes applicatifs -- par exemple, lors d'une démarche centrée utilisation (UCD) -- ou encore pour valider qu'une voie alternative (d'accomplissement d'une tâche) offre de meilleures performances qu'avec d'autres techniques d'interaction \cite{greenberg2008usability}. Les tâches testées sont alors de haut niveau, spécifiques à des applicatifs et identiques ou représentatives des tâches des personnes utilisatrices finales. Tester la conception n'implique pas toujours de devoir utiliser des prototypes opérationnels~: des maquettes de basse fidélité permettent de tester la conception avec des personnes utilisatrices dès les premières étapes. Trois illustrations de la littérature sont, par exemple~:
\begin{enumerate}[label=(E\arabic*)]\setcounter{enumi}{15}
\item Tester des indices d'invitation et de permission pour des portes à changement de forme \cite{economidou2021nodoor};
\item Tester les modes novices et experts pour une technique de menu en mesurant l'apprentissage, la précision et la performance \cite{francone2010wavelet};
\item Tester la rigueur d'exécution et la créativité d'expression d'une interface numérique de dessin par stylet et toucher \cite{xia2018dataink}.
\end{enumerate}

\subsubsection{Les études optimisant (ou explorant) la conception}  

Recourir à des études expérimentales pour optimiser ou explorer des possibilités de conception, permet de rationaliser la conception, de sorte à prévenir les ``\textit{fausses idées d'une intuition égocentrique}'' \cite{landauer1997behavioral} (cité dans \cite{hornbaek2013whys}). Trois illustrations de la littérature sont les suivantes~:
\begin{enumerate}[label=(E\arabic*)]\setcounter{enumi}{18}
\item Explorer les formes inférant le mieux les tâches de contrôle du trafic aérien \cite{gramlich2022atc};
\item Explorer les approches de séparation et d'intégration des degrés de liberté pour interagir avec des tablettes par stylet et toucher \cite{pfeuffer2021bi3d};
\item Explorer des menus textuels et pictographiques pour le guide d'une télévision \cite{westerink1998tvmenus}.
\end{enumerate}

\subsubsection{Les études comparant des conceptions} 

Comparer plusieurs conceptions alternatives est aussi possible, au travers des \textit{études expérimentales comparant des conceptions}. Trois illustrations de la littérature sont les suivantes~:
\begin{enumerate}[label=(E\arabic*)]\setcounter{enumi}{21}
\item Comparer l'efficacité de trois techniques de menu pour interagir en conservant les mains libres \cite[Étude expérimentale~2]{bailly2011freehand};
\item Comparer trois techniques de navigation pour des voyages virtuels à 360 degrés affichés dans des casques de réalité virtuelle \cite{li2021tapestries};
\item Comparer trois techniques d'entrées et deux de sorties pour des tâches de positionnement 3D dans des casques de réalité virtuelle \cite{sun2018input}.
\end{enumerate}

\subsection{Les études expérimentales multi-facettes}

Des études expérimentales peuvent combiner plusieurs facettes au sein d'un unique protocole, résolvant alors un \textit{problème mixte} \cite{oulasvirta2016solving}. Dans ce cas, une même recherche peut donc servir plusieurs questions à la fois.
Trois illustrations de la littérature, de ces \textit{études expérimentales mixtes} consistent, par exemple, à~:
\begin{enumerate}[label=(E\arabic*)]\setcounter{enumi}{24}
\item Explorer la posture de main préférée, en trouvant la meilleure position de caméra et en optimisant l'algorithme pour une technique de menu laissant les mains libres \cite[étude utilisatrice~1]{bailly2011freehand} (\cad{} une étude utilisatrice mixte \textit{optimisant la conception} et \textit{optimisant la technologie});
\item Tester un index d'extension du corps par des outils, basé sur les temps de réponse de stimuli, selon trois technologies \cite[Expérimentation~1]{bergstrom2019tool} (\cad{} une étude utilisatrice mixte \textit{testant une hypothèse} et \textit{comparant des technologies}) et selon trois conceptions \cite[Expérimentation~2]{bergstrom2019tool} (\cad{} une étude utilisatrice mixte \textit{testant une hypothèse} et \textit{comparant des conceptions});
\item Investiguer des mécanismes d'entrées 3D pour de grands afficheurs \cite{laundry2010input3d} (\cad{} une étude utilisatrice mixte \textit{testant la technologie} et \textit{comparant des technologies}).
\end{enumerate}

\subsection{Les études expérimentales lors de cas applicatifs}

Le présent cadre motivationnel nous permet, en particulier, d'examiner les études expérimentales conduites lors du traitement de cas applicatifs. Ces études expérimentales se révèlent être de natures extrêmement variées, pouvant combiner chacune des trois facettes, pour aboutir à des observations très diverses. Trois illustrations de la littérature, de ces \textit{études expérimentales de cas applicatifs} consistent, par exemple, à~:
\begin{enumerate}[label=(E\arabic*)]\setcounter{enumi}{27}
\item Tester l'influence des émotions sur les erreurs de saisie de nombres, dans les salles opératoires, au travers de deux applications logicielles \cite{cairns2014emotion} (\cad{} une étude utilisatrice mixte \textit{testant une hypothèse} et \textit{comparant des conceptions});
\item Comparer des versions tactile multi-point et tangible d'une interface numérique, pour l'exploration de données, dans des expositions muséales \cite{joyce2015versus} (\cad{} une étude utilisatrice \textit{comparant la technologie});
\item Comparer quatre conditions d'entrées et deux techniques d'interaction pour des salles de contrôle, selon le temps d'achèvement et la précision de remémoration de séquences d'actions de contrôle \cite{muller2018back} (\cad{} une étude utilisatrice mixte \textit{comparant la technologie} et \textit{comparant des conceptions}).

\end{enumerate}

Cependant, parmi les indices collectés au travers de toutes ces études expérimentales que nous venons de cartographier, le construit d'utilisabilité est souvent considéré comme la \g{preuve ultime}, alors qu'il apporte seulement une ``preuve d'existence'' \cite{greenberg2008usability}. Les suppositions émises, sur la supériorité d'une technologie ou d'un choix de conception et sur la validité d'une hypothèse de causalité, nécessitent de s'appuyer sur des preuves observées plus d'une seule et unique fois \cite{greenberg1992weak,greenberg2008usability,hornbaek2014once,hornbaek2015wrong}. Ainsi, nous ne devons pas nous contenter de confirmations ou de réfutations obtenues lors d'études expérimentales isolées \cite{hornbaek2015wrong}. Pour pallier ce constat, cet article propose dans la suite une méthode de recherche permettant un effort à l'échelle de communautés de recherche en IHM, de sorte à pouvoir générer et répliquer des résultats obtenus à partir d'un même format d'études expérimentales conduites sur des cas applicatifs, et qui pourront ensuite faire l'objet d'efforts de généralisation. Aussi, pour exprimer les conditions expérimentales de ces résultats, la section suivante rassemble des éléments de caractérisation des tâches et des prototypes expérimentaux.

\section{Une caractérisation des conditions expérimentales}\label{sec-conditions}

Cette section propose de caractériser les conditions expérimentales de sorte à pouvoir mieux décrire les contextes d'obtention de résultats empiriques. Les trois dimensions proposées sont le type de tâche, le type de prototype et le profil de personnes participantes. Le but est autant d'aider à exprimer les résultats empiriques et leur contexte d'apparition, que d'aider à préparer des protocoles expérimentaux.

De fait, les modèles de tâches s'appliquent principalement à décomposer les tâches en hiérarchies et en relations temporelles impliquées dans des systèmes et applications d'envergure. Même si ces outils deviennent surdimensionnés lors de la description de résultats empiriques obtenus sur un morceau d'interface numérique ou une sous-partie d'une application, les grands principes qu'ils développent restent applicables et peuvent, à moindre niveau de détail, se voir réemployés. Par exemple, une simplification commune à la description de nombreux protocoles amène le plus souvent à considérer deux niveaux, mêlant hiérarchie et cognition~: les tâches de bas niveau (TBN) et de haut niveau (THN); ainsi que des relations temporelles restreintes à des tâches à difficulté croissantes (TDC) et des tâches à difficulté alternante (TDA).

Toutefois, ces descriptions restent purement articulatoires ou cognitives. Préciser les cadres expérimentaux bénéficierait d'une description du prototype impliqué, c'est-à-dire du morceau d'interface numérique ou de la sous-partie d'application, et donc du cadre opératoire des tâches permises par ce prototype. Plus concrètement, l'objectif est de pouvoir décrire de manière plus fine l'espace des conditions expérimentales entre deux extrema, où d'un côté des tâches de bas de niveau sont impliquées sur des morceaux d'interfaces dans des expérimentations permettant d'obtenir des résultats le plus générique possible (\pex{} dans le but de vérifier une hypothèse de causalité), et où de l'autre côté impliquer des tâches de plus haut niveau permet de mieux évaluer l'utilisabilité finale d'une sous-partie d'une application (\pex{} dans le but de valider une conception). Un niveau de description pertinent serait, par exemple, de s'appuyer sur la portée applicative du prototype impliqué, comme déjà proposé pour discriminer des morceaux réutilisables d'interfaces tangibles avec les deux niveaux \textit{fondement} et \textit{domaine} \cite{ullmer2008core}. Les éléments tangibles de fondement répondent à des besoins communs à diverses applications (\cad{} au niveau des opérations), alors que les éléments tangibles de domaine répondent au besoin spécifique d'un domaine d'application (\cad{} au niveau applicatif). Cette façon de distinguer, selon l'étendue du besoin, semble pertinente pour discriminer les prototypes impliqués dans les protocoles expérimentaux, ainsi que les tâches qu'ils supportent. Ainsi, dans la suite, quatre niveaux de besoin sont retenus pour déterminer la nature des tâches et trois pour la portée des prototypes.

\subsection{Les tâches expérimentales}

Les tâches impliquées dans les protocoles expérimentaux sont discriminées selon les deux axes suivants~:

\begin{enumerate}

\item \textbf{La nature de la tâche} distingue quatre cas~:

\begin{enumerate}[label=(\roman*)]
\item \textit{Applicative}~: La tâche même des personnes utilisatrices finales (TUF), ou tâche finale, ou tâche cible, qui est exécutée par utilisation d'une application conçue pour répondre au besoin exprimé de manière spécifique. Cette tâche traite directement les données des personnes utilisatrices finales ou bien des échantillons.
\item \textit{Représentative}~: Une tâche représentative de la tâche des personnes utilisatrices finales (TRU), qui s'en inspire mais la simplifie ou l'abrège pour s'adapter aux contraintes du cadre expérimental (\pex{} durée des expérimentations, effort et coût de développement, possibilités de calcul). Cette tâche traite directement les données des personnes utilisatrices finales, ou bien des échantillons, ou encore un jeu de données alternatif reprenant les caractéristiques des données initiales mais simplifiant ou abrégeant l'exécution de la tâche.
\item \textit{Domaine}~: La tâche répond à un besoin commun à diverses applications (\pex{} sélectionner dans un menu, spécifier une couleur avec une palette, ou lire les valeurs d'un diagramme). Cette tâche traite des données choisies pour répondre aux besoins de l'expérimentation, selon la question de recherche soulevée.
\item \textit{Opération}~: La tâche est une opération, c'est-à-dire une tâche atomique ou élémentaire, qui ne se décompose pas en sous-tâches. Cette tâche ne dépend pas d'un domaine d'application en particulier ou se retrouve dans énormément d'interfaces humain-machine (\pex{} tâche de pointage). Cette tâche traite des données choisies pour répondre aux besoins de l'expérimentation, selon la question de recherche soulevée (\pex{} impliquant la répétition et cadrant la variation de la tâche de pointage \cite{iso:9241-9:2000}).
\end{enumerate}

\item \textbf{Le niveau de la tâche} dépend de trois facteurs~:

\begin{itemize}
\item \textit{Raisonnement}~: Résoudre la tâche implique-t-il un haut niveau de raisonnement de la part des personnes utilisatrices~?
\item \textit{Hiérarchie}~: La procédure de la tâche nécessite-t-elle d'accomplir des sous-tâches et de combiner leurs résultats pour résoudre un problème global~?
\item \textit{Données}~: Le jeu de données rend-t-il la tâche difficile~? (\pex{} par sa taille, par sa complexité, par le nombre d'étapes)
\end{itemize}

\end{enumerate}

Vérifier un seul des trois facteurs peut suffire à rendre une tâche difficile et à pouvoir la qualifier de haut niveau (THN). Lorsqu'aucun des trois facteurs n'est vérifié, la tâche sera de bas niveau (TBN). Lorsque déterminer le niveau de difficulté de la tâche est déterminant pour qu'une tâche soit représentative de la tâche des personnes utilisatrices finales, des questionnaires peuvent aider à mesurer et calibrer le niveau subjectif de charge mentale perçu \cite{kosch2023workload} (\pex{} \cite{hart1988tlx}).

\subsection{Les prototypes expérimentaux}

Les prototypes expérimentaux permettant l'accomplissement des tâches susmentionnées, ils peuvent être distingués en fonction du type des tâches impliquées. Pour ce faire, la dimension proposée est \textbf{la portée d'utilisation du prototype} avec les trois valeurs suivantes~:

\begin{enumerate}[label=(\roman*)]
\item Un \textit{prototype avec une portée applicative} permet d'accomplir la tâche des personnes utilisatrices finales (\pex{} le contrôle du trafic aérien \cite{mackay1999strips,vinot2014air}) ou une version représentative de cette tâche (\pex{} une alternance de tâches de difficultés variables en milieu stérile \cite{bailly2021exploration}). Néanmoins, un protocole expérimental peut parfaitement vouloir seulement mesurer une sous-tâche ou viser la mesure d'une opération impliquée dans l'exécution de la tâche.
\item Un \textit{prototype avec la portée d'un domaine} permet de répondre à un besoin commun à diverses applications (\pex{} des diagrammes à bâtons \cite{jansen2013evaluation,ren2021comparing}). Les tâches exécutées peuvent être spécifiques au domaine (comme la lecture bas niveau des informations \cite{jansen2013evaluation}) ou spécifiques au jeu de données (\pex{} répondre à des questions spécifiques au jeu de données en interprétant le diagramme \cite{ren2021comparing}).
\item Un \textit{prototype avec la portée d'une opération} est restreint à l'accomplissement de cette seule opération (\pex{} une tâche de pointage \cite{fitts1954law}). 
\end{enumerate}

Même si certains des concepts développés dans cet article peuvent s'appliquer aux deux autres types de prototypes, le seul cas considéré ici est celui des prototypes avec une portée applicative (sous la dénomination abrégée de \g{prototype applicatif}).

\subsection{Le profil de personnes participantes}

Le profil des personnes participantes, qui peut aussi bien servir l'hypothèse et l'expression du résultat (\pex{} en distinguant deux catégories de personnes utilisatrices), que fournir un indice de confiance dans la qualité des résultats empiriques, est discriminé selon trois axes~:

\begin{enumerate}
\item \textbf{Pratique} : Le niveau de pratique préalable du dispositif interaction, de la technologie utilisée, ou de la technique d'interaction (\pex{} \textit{novice}, \textit{débutant}, \textit{compétent}, \textit{performant} ou \textit{expert}).
\item \textbf{Spécialité} : Une personne \textit{spécialiste} est la cible même du besoin applicatif (\pex{} métier, loisir, client). Elle a normalement une connaissance du type des données et de l'objectif et de la procédure de la tâche à accomplir, sans pour autant connaître la technologie ou l'application utilisée. Être spécialiste et être expert sont donc deux choses distinctes. 
\item \textbf{Origine} : La façon dont la personne a été sélectionnée, que ce soit par exemple des \textit{collègues} de laboratoire ou de département (\pex{} personnels de recherche, d'enseignement, administratifs), des personnes \textit{lambdas} recrutées à la volée (\pex{} visiteurs d'un lieu, service de micro-travail en ligne, passants), ou encore la constitution d'un \textit{échantillon} représentatif d'une population.
\end{enumerate}

\section{La diffraction de boucles d'interaction}\label{sec-diffraction}

Cette section conceptualise un moyen de révélation des phénomènes interactionnels, en introduisant un outil imaginaire (par une métaphore en optique ondulatoire), avec le but de faciliter la compréhension dans la description des protocoles, le dialogue lors de la préparation d'un protocole, mais aussi l'interprétation des résultats.

\begin{figure}\centering
\def\svgwidth{8.47cm}
  {\small \begingroup \makeatletter \providecommand\color[2][]{\errmessage{(Inkscape) Color is used for the text in Inkscape, but the package 'color.sty' is not loaded}\renewcommand\color[2][]{}}\providecommand\transparent[1]{\errmessage{(Inkscape) Transparency is used (non-zero) for the text in Inkscape, but the package 'transparent.sty' is not loaded}\renewcommand\transparent[1]{}}\providecommand\rotatebox[2]{#2}\newcommand*\fsize{\dimexpr\f@size pt\relax}\newcommand*\lineheight[1]{\fontsize{\fsize}{#1\fsize}\selectfont}\ifx\svgwidth\undefined \setlength{\unitlength}{421.54304528bp}\ifx\svgscale\undefined \relax \else \setlength{\unitlength}{\unitlength * \real{\svgscale}}\fi \else \setlength{\unitlength}{\svgwidth}\fi \global\let\svgwidth\undefined \global\let\svgscale\undefined \makeatother \begin{picture}(1,0.3076355)\lineheight{1}\setlength\tabcolsep{0pt}\put(0,0){\includegraphics[width=\unitlength,page=1]{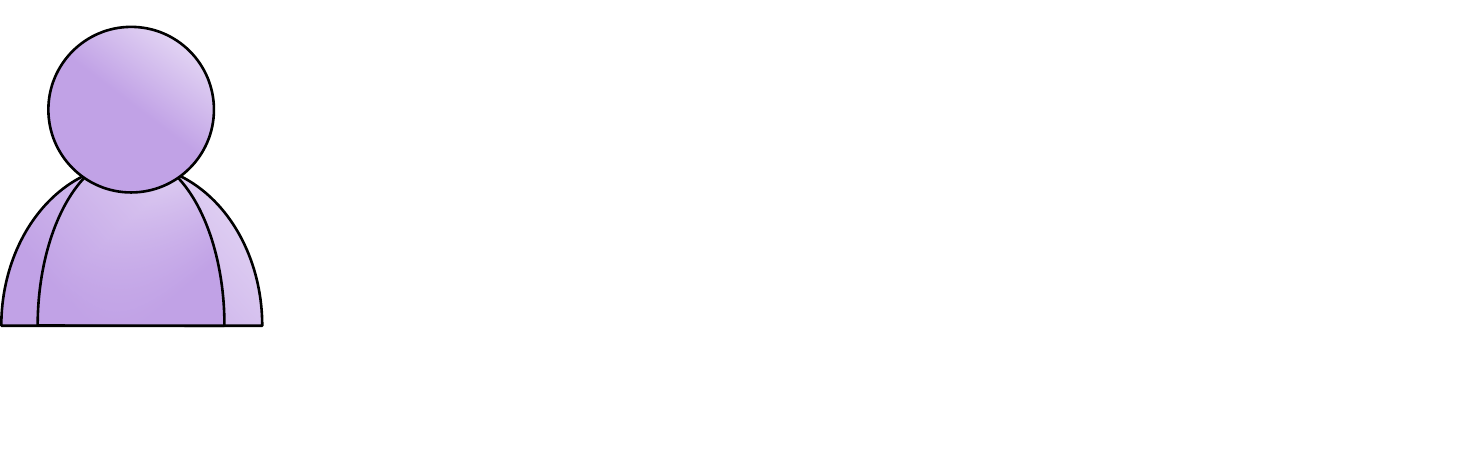}}\put(0.50566968,0.26839096){\color[rgb]{0,0,0}\makebox(0,0)[t]{\lineheight{1.25}\smash{\begin{tabular}[t]{c}\textbf{Sortie}\end{tabular}}}}\put(0.50566968,0.07623958){\color[rgb]{0,0,0}\makebox(0,0)[t]{\lineheight{1.25}\smash{\begin{tabular}[t]{c}\textbf{Entrée}\end{tabular}}}}\put(0.50566968,0.16946326){\color[rgb]{0,0,0}\makebox(0,0)[t]{\lineheight{1.25}\smash{\begin{tabular}[t]{c}\textbf{BOUCLE D'INTERACTION}\end{tabular}}}}\put(0,0){\includegraphics[width=\unitlength,page=2]{interaction_loop-fr.pdf}}\put(0.91566446,0.06079867){\color[rgb]{0,0,0}\makebox(0,0)[t]{\lineheight{1.10000002}\smash{\begin{tabular}[t]{c}Système\\interactif\end{tabular}}}}\put(0.08873005,0.03700275){\color[rgb]{0,0,0}\makebox(0,0)[t]{\lineheight{1.10000002}\smash{\begin{tabular}[t]{c}Personne\\utilisatrice\end{tabular}}}}\put(0,0){\includegraphics[width=\unitlength,page=3]{interaction_loop-fr.pdf}}\end{picture}\endgroup  }
  \caption{Les deux sens de communication des boucles d'interaction. L'interaction peut se voir comme un dialogue composé d'échanges cycliques qui empruntent des boucles d'interaction.}
\Description{A user is represented on the left. An interactive system is represented on the right. Two arrows create an ``interaction loop'' that goes from the user to the interactive system (\ie{} input) and from the interactive system to the user (\ie{} output).}
  \label{fig:loops}
\end{figure}

\begin{figure}\centering
  \begin{subfigure}[b]{\linescale\linewidth}
    \centering
\def\svgwidth{8.47cm}
    {\small \begingroup \makeatletter \providecommand\color[2][]{\errmessage{(Inkscape) Color is used for the text in Inkscape, but the package 'color.sty' is not loaded}\renewcommand\color[2][]{}}\providecommand\transparent[1]{\errmessage{(Inkscape) Transparency is used (non-zero) for the text in Inkscape, but the package 'transparent.sty' is not loaded}\renewcommand\transparent[1]{}}\providecommand\rotatebox[2]{#2}\newcommand*\fsize{\dimexpr\f@size pt\relax}\newcommand*\lineheight[1]{\fontsize{\fsize}{#1\fsize}\selectfont}\ifx\svgwidth\undefined \setlength{\unitlength}{421.54304528bp}\ifx\svgscale\undefined \relax \else \setlength{\unitlength}{\unitlength * \real{\svgscale}}\fi \else \setlength{\unitlength}{\svgwidth}\fi \global\let\svgwidth\undefined \global\let\svgscale\undefined \makeatother \begin{picture}(1,0.3076355)\lineheight{1}\setlength\tabcolsep{0pt}\put(0,0){\includegraphics[width=\unitlength,page=1]{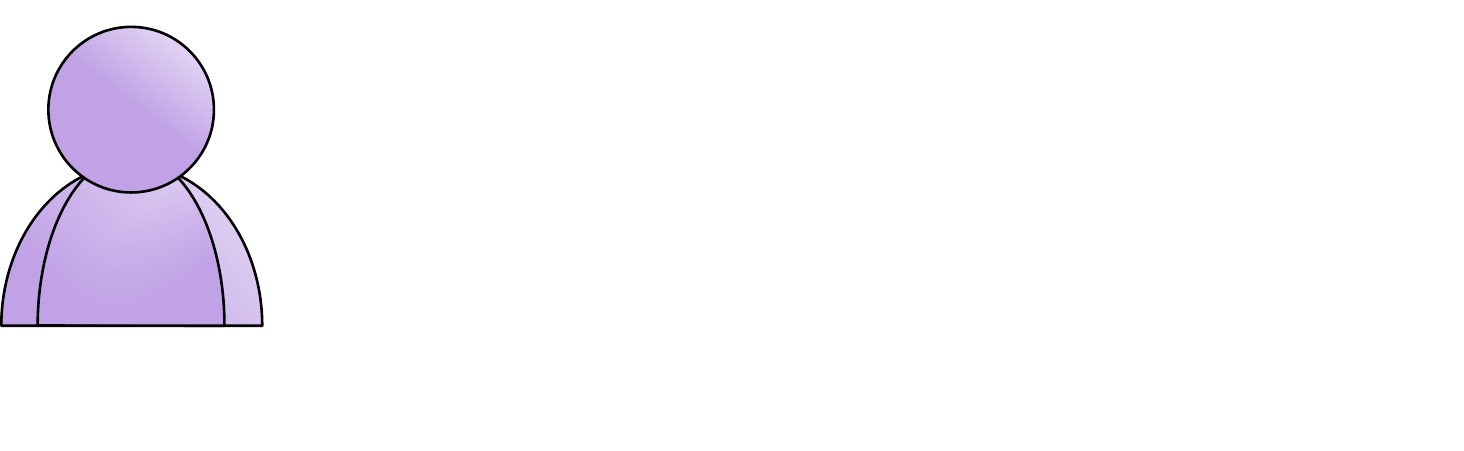}}\put(0.91566435,0.06079867){\color[rgb]{0,0,0}\makebox(0,0)[t]{\lineheight{1.10000002}\smash{\begin{tabular}[t]{c}Système\\interactif\end{tabular}}}}\put(0,0){\includegraphics[width=\unitlength,page=2]{diffracted_input-fr.pdf}}\put(0.5071129,0.09728679){\color[rgb]{0,0,0}\makebox(0,0)[t]{\lineheight{1.45000005}\smash{\begin{tabular}[t]{c}\textit{e1}\\\textit{e2}\\\textit{e3}\end{tabular}}}}\put(0,0){\includegraphics[width=\unitlength,page=3]{diffracted_input-fr.pdf}}\put(0.50136495,0.13307835){\color[rgb]{0,0,0}\makebox(0,0)[t]{\lineheight{1.25}\smash{\begin{tabular}[t]{c}\textbf{Entrée diffractée}\end{tabular}}}}\put(0.50566968,0.26839096){\color[rgb]{0,0,0}\makebox(0,0)[t]{\lineheight{1.25}\smash{\begin{tabular}[t]{c}\textbf{Sortie}\end{tabular}}}}\put(0,0){\includegraphics[width=\unitlength,page=4]{diffracted_input-fr.pdf}}\put(0.08873005,0.03700275){\color[rgb]{0,0,0}\makebox(0,0)[t]{\lineheight{1.10000002}\smash{\begin{tabular}[t]{c}Personne\\utilisatrice\end{tabular}}}}\put(0,0){\includegraphics[width=\unitlength,page=5]{diffracted_input-fr.pdf}}\end{picture}\endgroup  }
    \caption{Diffraction en entrée.}
    \label{fig:loops-prims-input}
    \Description{A user is represented on the left. An interactive system is represented on the right. An arrow from the user to the interactive system is diffracted between two prisms (\ie{} diffracted input), thus creating three input diversions i1, i2, and i3. Another arrow goes from the interactive system to the user (\ie{} output).}
  \end{subfigure}
  \begin{subfigure}[b]{\linescale\linewidth}
    \centering
    \vspace{12pt}
\def\svgwidth{8.47cm}
    {\small \begingroup \makeatletter \providecommand\color[2][]{\errmessage{(Inkscape) Color is used for the text in Inkscape, but the package 'color.sty' is not loaded}\renewcommand\color[2][]{}}\providecommand\transparent[1]{\errmessage{(Inkscape) Transparency is used (non-zero) for the text in Inkscape, but the package 'transparent.sty' is not loaded}\renewcommand\transparent[1]{}}\providecommand\rotatebox[2]{#2}\newcommand*\fsize{\dimexpr\f@size pt\relax}\newcommand*\lineheight[1]{\fontsize{\fsize}{#1\fsize}\selectfont}\ifx\svgwidth\undefined \setlength{\unitlength}{421.54304528bp}\ifx\svgscale\undefined \relax \else \setlength{\unitlength}{\unitlength * \real{\svgscale}}\fi \else \setlength{\unitlength}{\svgwidth}\fi \global\let\svgwidth\undefined \global\let\svgscale\undefined \makeatother \begin{picture}(1,0.34590758)\lineheight{1}\setlength\tabcolsep{0pt}\put(0,0){\includegraphics[width=\unitlength,page=1]{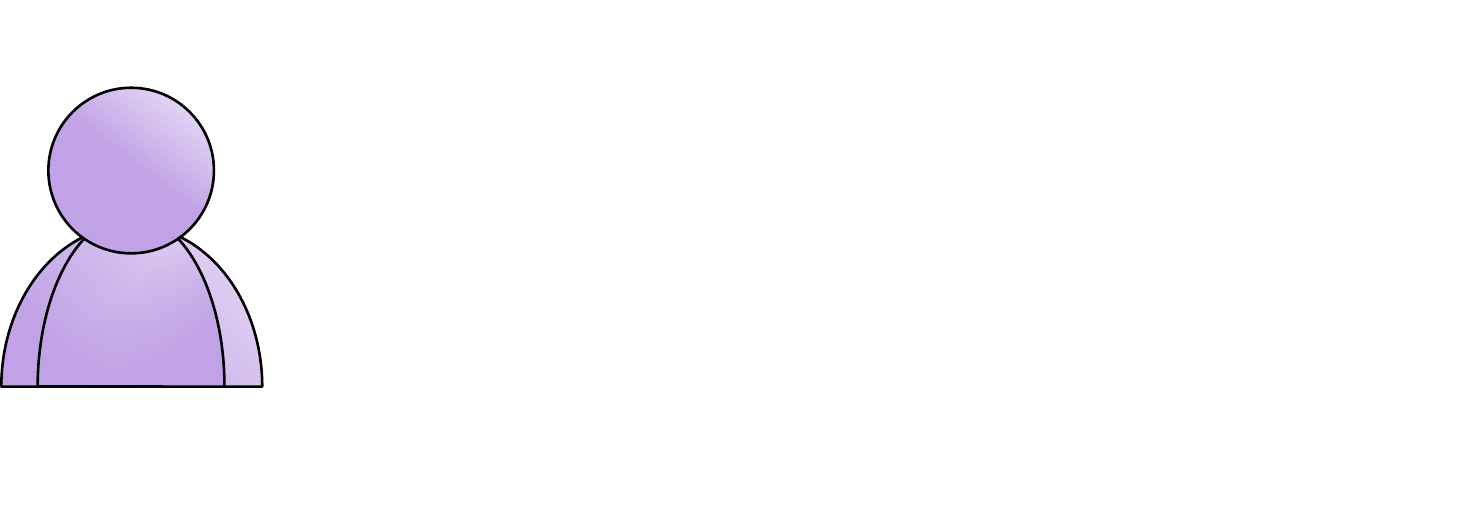}}\put(0.91566446,0.05747751){\color[rgb]{0,0,0}\makebox(0,0)[t]{\lineheight{1.10000002}\smash{\begin{tabular}[t]{c}Système\\interactif\end{tabular}}}}\put(0,0){\includegraphics[width=\unitlength,page=2]{diffracted_output-fr.pdf}}\put(0.50711284,0.32169639){\color[rgb]{0,0,0}\makebox(0,0)[t]{\lineheight{1.45000005}\smash{\begin{tabular}[t]{c}\textit{s1}\\\textit{s2}\\\textit{s3}\end{tabular}}}}\put(0,0){\includegraphics[width=\unitlength,page=3]{diffracted_output-fr.pdf}}\put(0.5013649,0.19024512){\color[rgb]{0,0,0}\makebox(0,0)[t]{\lineheight{1.25}\smash{\begin{tabular}[t]{c}\textbf{Sortie diffractée}\end{tabular}}}}\put(0.50566968,0.07291842){\color[rgb]{0,0,0}\makebox(0,0)[t]{\lineheight{1.25}\smash{\begin{tabular}[t]{c}\textbf{Entrée}\end{tabular}}}}\put(0,0){\includegraphics[width=\unitlength,page=4]{diffracted_output-fr.pdf}}\put(0.08873005,0.03368159){\color[rgb]{0,0,0}\makebox(0,0)[t]{\smash{\begin{tabular}[t]{c}Personne\\utilisatrice\end{tabular}}}}\put(0,0){\includegraphics[width=\unitlength,page=5]{diffracted_output-fr.pdf}}\end{picture}\endgroup  }
    \caption{Diffraction en sortie.}
    \label{fig:loops-prims-output}
    \Description{A user is represented on the left. An interactive system is represented on the right. Another arrow goes from the user to the interactive system (\ie{} input). An arrow from the interactive system to the user is diffracted between two prisms (\ie{} diffracted output), thus creating three output diversions o1, o2, and o3. }
  \end{subfigure}
  \caption{Les prismes interactionnels sont des outils imaginaires permettant de diffracter les boucles d'interaction. Diffracter en un unique point permet de révéler une propriété interactionnelle.}
\label{fig:loops-prims}
\end{figure}

Cet article considère l'interaction comme un dialogue et prend comme moyen d'observation de l'interaction les incidences lors de l'altération des boucles d'interaction impliquées dans ce dialogue. Les boucles d'interaction sont constituées des entrées de la personne utilisatrice vers le système interactif et des sorties du système interactif vers la personne utilisatrice (voir \autoref{fig:loops}). L'altération (ou déviation, ou révélation) des boucles d'interaction s'effectue par la variation d'un paramètre du système interactif en un point d'une boucle. L'observation des incidences de cette altération peut éventuellement intervenir lors de l'exécution de plusieurs tâches de niveau de difficulté variable. Faire varier un paramètre revient à diffracter une boucle d'interaction en un point, à l'aide d'un outil imaginaire appelé \g{prisme interactionnel}, en lui faisant prendre plusieurs valeurs possibles (\cad{} plusieurs voies). Un prisme diffracte soit les entrées vers le système (voir \autoref{fig:loops-prims-input}), soit les sorties vers la personne utilisatrice (voir \autoref{fig:loops-prims-output}). Un paramètre peut, par exemple, caractériser certains aspects de l'interface numérique comme une vitesse, une taille, une couleur, une forme, un poids, une texture, un multiplexage ou une position. Ces paramètres doivent être impliqués dans une boucle d'interaction sollicitée et déterminante lors de l'exécution de la tâche, de sorte que leur variation influe sur la boucle d'interaction.

D'un point de vue statistique, un prisme interactionnel décrit simplement une variable indépendante. En revanche, le vocabulaire de prisme interactionnel se veut spécifique aux boucles d'interaction. De plus, cette métaphore permet une représentation graphique des phénomènes observés et leur mise en rapport avec les boucles d'interaction impliquées dans une interface en particulier.

Ainsi, la diffraction de boucles d'interaction permet d'imposer une certaine prise de vue~: observer l'interaction plutôt que les interfaces. Pour ce faire, l'observation de l'interaction par diffraction est simplement une façon de procéder à une étude utilisatrice (variation des valeurs d'un paramètre) et une façon d'en interpréter les résultats (en considérant avoir modifié une boucle). L'idée est, comme pour l'étude de rayons invisibles en optique ondulatoire, de rendre apparentes les boucles d'interaction du dialogue par leur diffraction en un point. Les incidences résultent donc de la diffraction en ce point, rendant ainsi la diffraction visible, à condition que le paramètre et les valeurs choisies soient suffisamment impactantes. La diffraction d'une boucle en de multiples points à la fois (\pex{} en entrée et en sortie), ou de multiples boucles en même temps, rendrait difficile l'interprétation des résultats sans connaissance préalable de l'effet individuel de l'altération en chacun de ces points. Ainsi, de sorte à pouvoir cumuler les résultats obtenus par diffraction de boucles, la section suivante introduit les propriétés interactionnelles.

\section{Les propriétés interactionnelles}\label{sec-properties}

Nous définissons les \g{propriétés interactionnelles} comme des assertions pouvant être vraies ou fausses (au même sens que les propriétés en mathématiques) concernant l'effet d'un paramètre donné sur l'interaction. Typiquement, les propriétés interactionnelles expliquent comment les caractéristiques des interfaces humain-machine affectent l'interaction à cause des capacités humaines. Ces propriétés sont révélées par diffraction de boucles d'interaction, en faisant varier les valeurs d'un paramètre et, éventuellement, le niveau de difficulté des tâches. Une présentation particulière des propriétés interactionnelles permet de les exprimer et de les interpréter selon un cadre d'étude de l'interaction, vue comme un dialogue composé d'échanges cycliques. Pour ce faire, cette section propose un formalisme de description des propriétés interactionnelles. Ce formalisme a autant pour but de conditionner l'observation des propriétés interactionnelles, que d'en constituer un cadre d'interprétation, que d'en alimenter un catalogue utilisable en recherche et en ingénierie indépendemment des prototypes et des technologies mises en \oe{}uvre.

\subsection{Formalisme}

Une propriété interactionnelle est le résultat d'une expérimentation qui se décrit par une expression et de brèves informations contextuelles ($\bullet$~: nécessaire, $\circ$~: facultatif)~:

\begin{itemize}
\item \textbf{Prisme}~: Le type de prisme utilisé pour révéler le paramètre (entrée ou sortie).
\item \textbf{Boucle}~: Le nom de la boucle diffractée par le prisme.
\item \textbf{Portée du prototype}~: Le prototype impliqué dans la mesure (applicative, domaine, ou opération).
\item \textbf{Paramètre}~: Le paramètre impliqué dans l'interaction.
\item \textbf{Valeurs}~: Les valeurs concernées pour le paramètre.
\item [$\circ$] \textbf{Apparatus}~: L'appareillage et les dispositifs utilisés lors des observations.
\item [$\circ$] \textbf{Personnes participantes}~: Le nombre et de profil des personnes participantes.
\item \textbf{Tâches}~: La ou les tâches sur lesquelles la mesure des valeurs est effectuée (niveau et type de chaque tâche) et éventuellement la ou les autres tâches exécutées en plus (\pex{} tâche de distraction).
\item \textbf{Faculté humaine}~: Quel facteur limitant ou facilitant est stimulé par le paramètre et la tâche principale~?
\item \textbf{Propriété}~: Une expression logique indiquant les rapports entre les valeurs du paramètre (en précisant éventuellement si la difficulté de tâche entre en compte).
\item [$\circ$] \textbf{Théorie sous-jacente}~: Quelle théorie permet d'expliquer le résultat observé~? (si possible, préciser les références bibliographiques de la théorie concernée)
\end{itemize}

Les informations obligatoires sont nécessaires à la description de la propriété, alors que les informations facultatives sont données si elles sont utiles pour préciser la propriété observée (\pex{} apparatus), si elles sont pertinentes (\pex{} tâche de distraction), ou si elles sont connues (\pex{} théorie). La description faîte doit rester une brève synthèse des éléments déjà présents dans l'article. Les informations collectées doivent donc rester concises.

\subsection{Les études de propriétés}

Révéler une propriété interactionnelle s'obtient en isolant l'effet d'un paramètre, par diffraction d'une boucle d'interaction en un unique point. Différentes valeurs sont alors prises par un paramètre impliqué dans une boucle, et le niveau de tâche peut également varier pour évaluer son éventuelle implication dans la propriété interactionnelle. Cette approche résulte en un type particulier d'études expérimentales à deux facettes \textit{prospectant des causalités} et \textit{optimisant la conception} ou \textit{la technologie}, désigné comme des \ita{études de propriétés}. La section suivante propose une approche opportuniste, d'observation de propriétés interactionnelles, lors du traitement de cas applicatifs.

\section{Observer l'interaction en optimisant des prototypes applicatifs}\label{sec-proposition}

Cette section introduit comment préparer des études de propriétés interactionnelles, de sorte à pouvoir à la fois optimiser un prototype applicatif, tout en recueillant des observations permettant d'étayer des théories. La raison de cette approche duale est de pouvoir impulser et soutenir un élan, à l'échelle des communautés de recherche en IHM, permettant de générer et, surtout, de répliquer des résultats empiriques. Pouvoir combiner des approches au sein d'une même recherche (en l'occurrence d'optimisation et de réplication), est une démarche déjà acceptée et éprouvée en recherche mixte \cite{johnson2004mixed}. De plus, combiner des questions de conception et de causalité au sein d'une même étude utilisatrice est déjà admis comme étant problème mixte \cite{oulasvirta2016solving}. Les deux objectifs de traiter un cas d'étude et de d'explorer des propriétés peuvent donc fusionner au sein d'une même étude et d'une même recherche.

\begin{figure}\centering

  \def\titreA{\footnotesize{\textbf{1. Conception du prototype applicatif  (p.ex., centrée utilisation)}}}
  \def\titreB{\footnotesize{\textbf{2. Optimisation du prototype applicatif}}}
  \def\titreC{\footnotesize{\textbf{3. Comparaison avec d'autres technologies et conditions de base}}}
  \def\tit{\scriptsize}
  \def\sstit{\tiny}
  \def\svgwidth{7.5cm} {\small \begingroup \makeatletter \providecommand\color[2][]{\errmessage{(Inkscape) Color is used for the text in Inkscape, but the package 'color.sty' is not loaded}\renewcommand\color[2][]{}}\providecommand\transparent[1]{\errmessage{(Inkscape) Transparency is used (non-zero) for the text in Inkscape, but the package 'transparent.sty' is not loaded}\renewcommand\transparent[1]{}}\providecommand\rotatebox[2]{#2}\newcommand*\fsize{\dimexpr\f@size pt\relax}\newcommand*\lineheight[1]{\fontsize{\fsize}{#1\fsize}\selectfont}\ifx\svgwidth\undefined \setlength{\unitlength}{503.27393298bp}\ifx\svgscale\undefined \relax \else \setlength{\unitlength}{\unitlength * \real{\svgscale}}\fi \else \setlength{\unitlength}{\svgwidth}\fi \global\let\svgwidth\undefined \global\let\svgscale\undefined \makeatother \begin{picture}(1,1.18008211)\lineheight{1}\setlength\tabcolsep{0pt}\put(0,0){\includegraphics[width=\unitlength,page=1]{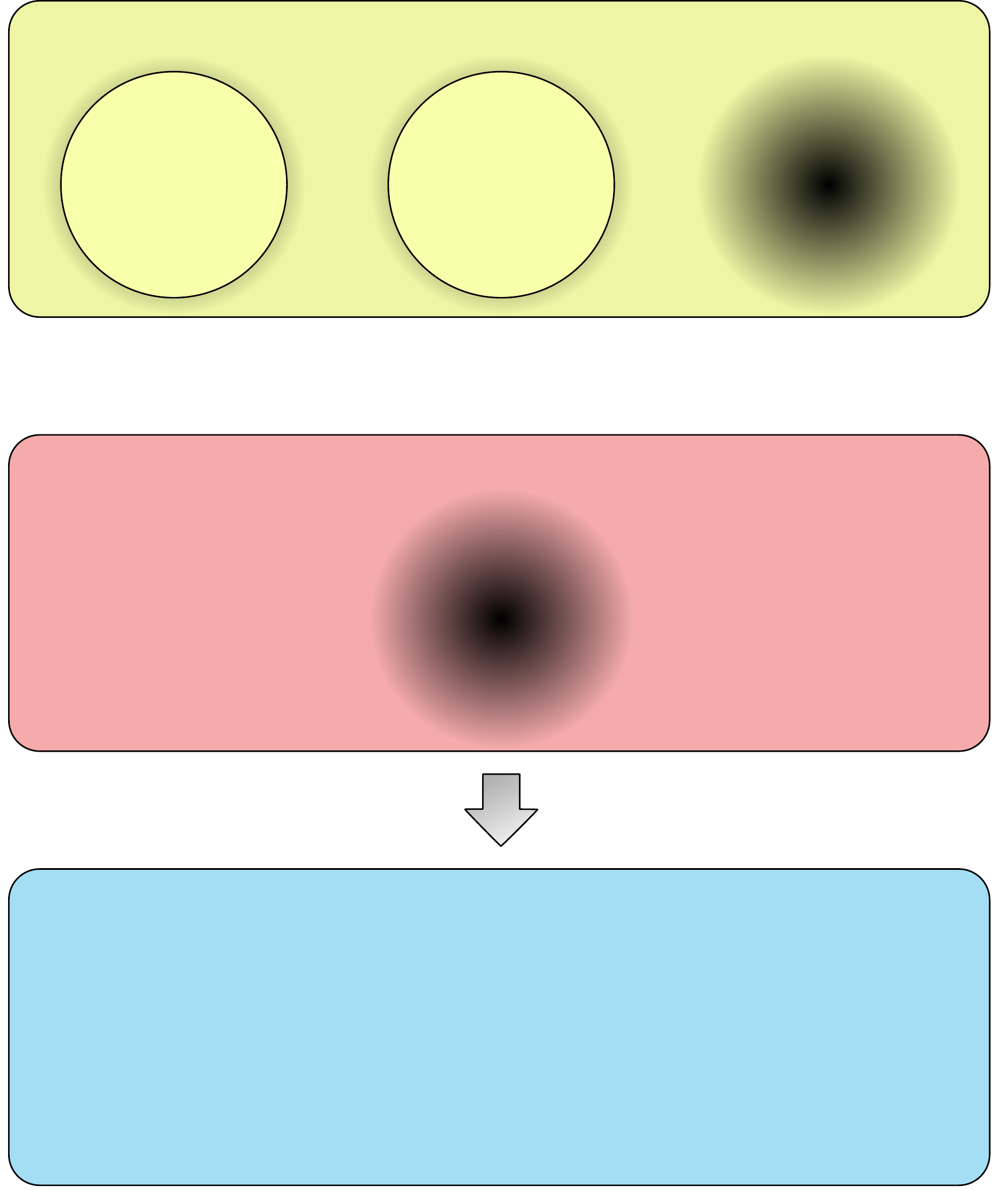}}\put(0.49779001,1.00820488){\color[rgb]{0,0,0}\makebox(0,0)[t]{\lineheight{0.94999999}\smash{\begin{tabular}[t]{c}\tit{Développement}\\\tit{du prototype}\end{tabular}}}}\put(0,0){\includegraphics[width=\unitlength,page=2]{workflow-fr.pdf}}\put(0.17245667,0.61205505){\color[rgb]{0,0,0}\makebox(0,0)[t]{\lineheight{0.94999999}\smash{\begin{tabular}[t]{c}\tit{Choix du}\\\tit{paramètre et}\\\tit{des valeurs}\\\tit{variantes}\end{tabular}}}}\put(0,0){\includegraphics[width=\unitlength,page=3]{workflow-fr.pdf}}\put(0.49779001,0.58721262){\color[rgb]{0,0,0}\makebox(0,0)[t]{\lineheight{0.94999999}\smash{\begin{tabular}[t]{c}\tit{Développement}\\\tit{d'extensions}\\\tit{au prototype}\end{tabular}}}}\put(0,0){\includegraphics[width=\unitlength,page=4]{workflow-fr.pdf}}\put(0.02282802,0.70466437){\color[rgb]{0,0,0}\makebox(0,0)[lt]{\lineheight{1.25}\smash{\begin{tabular}[t]{l}\titreB{}\end{tabular}}}}\put(0,0){\includegraphics[width=\unitlength,page=5]{workflow-fr.pdf}}\put(0.17291764,0.15322084){\color[rgb]{0,0,0}\makebox(0,0)[t]{\lineheight{0.94999999}\smash{\begin{tabular}[t]{c}\tit{Développement}\\\tit{de prototypes}\\\tit{concurrents}\end{tabular}}}}\put(0,0){\includegraphics[width=\unitlength,page=6]{workflow-fr.pdf}}\put(0.8219948,0.19138564){\color[rgb]{0,0,0}\makebox(0,0)[t]{\lineheight{0.94999999}\smash{\begin{tabular}[t]{c}\tit{Justification}\\\tit{du choix de}\\\tit{technologie}\\\tit{pour le cas}\\\tit{applicatif}\end{tabular}}}}\put(0.02307452,0.27393478){\color[rgb]{0,0,0}\makebox(0,0)[lt]{\lineheight{1.25}\smash{\begin{tabular}[t]{l}\titreC{}\end{tabular}}}}\put(0.49995085,0.20215525){\color[rgb]{0,0,0}\makebox(0,0)[t]{\lineheight{0.94999999}\smash{\begin{tabular}[t]{c}\tit{Étude}\\\tit{comparant}\\\tit{la technologie}\\\sstit{(Tâches de bas ou}\\\sstit{haut niveau)}\end{tabular}}}}\put(0.82063672,0.61511989){\color[rgb]{0,0,0}\makebox(0,0)[t]{\lineheight{0.94999999}\smash{\begin{tabular}[t]{c}\tit{Étude de}\\\tit{propriété}\\\sstit{(Tâches de bas}\\\sstit{et haut niveau)}\end{tabular}}}}\put(0.82207524,1.06442677){\color[rgb]{0,0,0}\makebox(0,0)[t]{\lineheight{0.94999999}\smash{\begin{tabular}[t]{c}\tit{Étude}\\\tit{testant la}\\\tit{conception}\\\sstit{(Tâches de bas ou}\\\sstit{haut niveau)}\end{tabular}}}}\put(0.17047862,1.02684285){\color[rgb]{0,0,0}\makebox(0,0)[t]{\lineheight{0.94999999}\smash{\begin{tabular}[t]{c}\tit{Cas applicatif}\\\sstit{(Recueil}\\\sstit{du besoin)}\end{tabular}}}}\put(0.02222398,1.1353432){\color[rgb]{0,0,0}\makebox(0,0)[lt]{\lineheight{1.25}\smash{\begin{tabular}[t]{l}\titreA{}\end{tabular}}}}\put(0,0){\includegraphics[width=\unitlength,page=7]{workflow-fr.pdf}}\end{picture}\endgroup  }
  \caption{Un exemple de nouvelle chaîne de traitement d'un cas applicatif, entée d'une phase d'optimisation. Optimiser le prototype avant comparaison paraît mieux indiqué, mais toute autre étape ou phase, et ordre de traitement, sont à adapter selon chaque situation.}
  \Description{The workflow comprises three successive phases (represented by three square blocks), each containing three successive steps (represented by three circles inside the blocks). Phase 1 -- Designing case study prototype (e.g., User-Centered Design): (a) a case study (requirements gathering) leads to (b) prototype development, which leads to (c) a design-testing user study (using low or high tasks). Phase 2 -- Optimizing case study prototype: (a) varied parameter and value choices lead to (b) prototype extensions' development, which leads to (c) a propriety study (using low and high tasks). Phase 3 -- Comparing with best-of-breed and previous baselines: (a) competitor prototype development leads to (b) a technology-comparing user study (using low or high tasks), which leads to (c) choice's justification for the Case Study.}
  \label{fig:workflow}
\end{figure}

L'approche proposée ajoute de nouvelles étapes lors du traitement de cas applicatifs, qui optimisent le prototype applicatif en étudiant une propriété interactionnelle, de sorte à étendre leur but initial de démontrer l'utilité d'une technologie. En plus des études expérimentales comparatives, communément déployées pour démontrer qu'une technologie est mieux adaptée que d'autres pour le développement d'un prototype applicatif, l'ajout d'une étude utilisatrice d'optimisation permet de trouver les meilleurs réglages pour une technologie et une conception donnée. La \autoref{fig:workflow} récapitule ces nouvelles étapes dans un exemple de traitement de cas applicatifs. Cependant, ces étapes de traitement sont à adapter aux spécificités de chaque cas. En effet, optimiser une nouvelle technologie ou une conception avant d'en faire la comparaison paraîtra mieux indiqué dans certains cas, alors que d'autres cas préféreront comparer des prototypes avant d'optimiser la solution retenue. En outre, d'autres études peuvent s'ajouter, comme des études pilotes, des études d'utilisabilité ou des études de terrains.

\begin{figure}\centering
  \includegraphics[trim={1.6cm 2.0cm 1.6cm 1.3cm},clip,width=8.47cm]{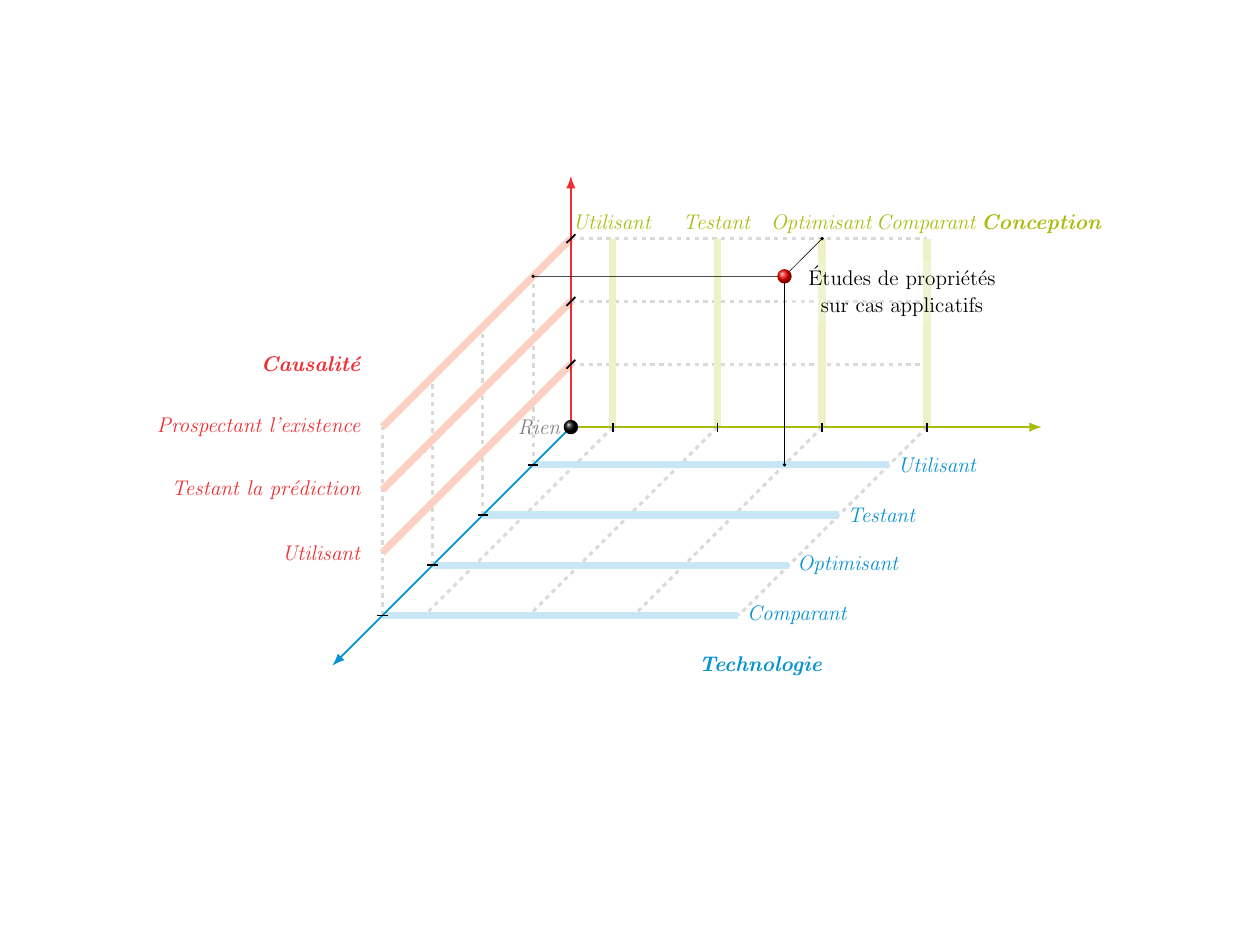} \caption{Les études de propriétés sur cas applicatifs au sein du cadre motivationnel des études expérimentales. Ces études à deux facettes, \textit{optimisant la conception} de prototypes applicatifs et \textit{prospectant des causalités}, utilisent aussi une technologie.}
  \Description{Propriety studies are projected on the three axes of the user study framework: using Technology, hunting Causality existence, and optimizing Design.}
  \label{fig:studies-framework-proprieties}
\end{figure}

En revanche, pour que les résultats puissent survivre aux prototypes, aux technologies et aux tâches impliquées, et se prêter facilement à la réplication, cette étude utilisatrice d'optimisation prend la forme d'une étude de propriété interactionnelle, en diffractant une boucle d'interaction par la variation des valeurs d'un unique paramètre. Les \ita{études de propriétés sur cas applicatifs} sont des études à deux facettes, \textit{optimisant la conception} de prototypes applicatifs et \textit{prospectant des causalités} (voir \autoref{fig:studies-framework-proprieties}). Pour ce faire, les études de propriétés reposent sur la variation de deux dimensions~: celle des paramètres liés à la conception du prototype applicatif et, éventuellement, le niveau de difficulté des tâches. La particularité de ces études repose sur la variation d'un seul et unique paramètre à la fois. Cette restriction permet de calibrer le format de ces études, de sorte à pouvoir opérer la prospection de causalités, au travers d'une diversité de cas applicatifs et de tâches applicatives. De la sorte, les observations calibrées ainsi recueillies peuvent être analysées comme des propriétés interactionnelles, pouvant aussi bien avoir répliqué une observation antérieure, que faire l'objet d'une réplication ultérieure.

Les observations effectuées informent alors sur la meilleure valeur à prendre pour le paramètre choisi (pour une tâche donnée, en sollicitant une certaine faculté humaine), contribuant ainsi à la fois à optimiser le prototype applicatif, mais aussi à établir une propriété interactionnelle (\cad{} en générant une première observation ou en répliquant une observation antérieure).

Aussi, comme le choix des paramètres variants sera postérieur au choix du cas applicatif, la propriété interactionnelle étayée sera connue tardivement. De même, la théorie sous-jacente associée à la propriété interactionnelle sera connue tardivement (et seulement dans le cas où une théorie préexiste). Établir ou consolider des propriétés interactionnelles, par réplication de diverses études de propriétés, demandera donc de changer la façon de penser en IHM. À ce dessein, la section suivante inscrit la présente approche dans la sérendipité.

\subsection{Choisir un paramètre à optimiser}\label{sec-serendipity-hypotheses}

Les études de propriétés nécessitent un paramètre à faire varier et de ce paramètre dépend la propriété qui sera observée. Or, le choix de ce paramètre doit se faire dans le seul et unique but d'optimiser l'accomplissement de la tâche du cas applicatif~: altérer un cas applicatif pour pouvoir répliquer une étude antérieure ou pour tester une théorie préexistante est à proscrire. En revanche, le paramètre choisi doit être suffisamment critique pour l'interaction et avoir un potentiel d'amélioration suffisant pour qu'une optimisation soit possible. Un ensemble de valeurs doit alors être défini pour ce paramètre (\pex{} de deux à quatre valeurs devraient permettre une inspection raisonnable et suffisante dans la plupart des cas). Les paramètres et les valeurs qui devraient influer sur l'interaction peuvent, par exemple, être déterminées par ce qu'elles impliquent au niveau des capacités humaines (\pex{} les capacités motrices, visuelles et auditives \citetext{\citealp[Ch.~1]{dix2003hci}}, les facultés de cognition, de raisonnement et de résolution de problèmes \citetext{\citealp[Ch.~1]{dix2003hci},\citealp[Ch.~4]{sharp2019beyond}} ou encore les aptitudes à collaborer, à coopérer et à vivre ensemble \citetext{\citealp[Ch.~5]{sharp2019beyond}}). Sinon, une brève étude pilote peut éventuellement aider à désigner le paramètre. Enfin, le choix du paramètre et des valeurs peut reposer sur l'intuition ou la connaissance des expérimentateurs, ou de son implication dans des études antérieures ou des propriétés interactionnelles existantes, mais surtout, ce choix doit encourager les recherches en IHM à dépasser l'attente de lois, de règles ou de théories préétablies, qui leur permettraient de choisir confortablement des paramètres et les valeurs à leur assigner (\cad{} sans prise de risque), en faisant confiance à la sérendipité.

Conduire des études de propriétés sur des cas applicatifs vient donc comme un palliatif pour redécouvrir la sérendipité, créant un cadre servant le développement et l'illustration des technologies, tout en offrant des conditions empiriques inhabituelles~: d'une part, les prototypes applicatifs et les tâches applicatives peuvent conduire à des observations inattendues aboutissant à la sérendipité \g{dirigée par l'observateur} et, d'autre part, le choix des paramètres et des valeurs qui varient peut provoquer de la sérendipité Walpolienne (\cad{} problème inattendu) et Mertonienne (\cad{} voie inattendue). De la sorte, la présente contribution évitera de produire des recommandations approfondies sur la façon de choisir des paramètres et des valeurs.

\subsection{Varier la difficulté des tâches}

\def\tasksfootnotetext{Les niveaux de tâche sont ici discriminés en adoptant les définitions telles que synthétisées par Balbo \cite{balbo1994thesis}. Dans la hiérarchie des tâches, les tâches composées requièrent l'accomplissement de sous-tâches, jusqu'à aboutir à des tâches élémentaires qui sont accomplies par exécutions d'actions (\cad{} du point de vue du système interactif) ou qui impliquent très peu de ressources cognitives (\cad{} du point de vue de la cognition, des tâches simples n'impliquant pas de superviser la progression d'un plan avec des branchements ou des itérations \cite{payne1986task}). La description des tâches dépend alors de la granularité prises pour les actions du système ou du niveau de compétence choisi pour les tâches élémentaires (qui dépendent du niveau d'expertise ou d'apprentissage des personnes utilisatrices). En effet, la complexité d'accomplissement d'une tâche varie en fonction du niveau d'expertise des personnes utilisatrices (par effet d'apprentissage)~: les tâches composées requièrent moins d'efforts cognitifs de la part des personnes expertes \cite{payne1986task}.}

Outre la variation d'un paramètre, les études de propriétés peuvent éventuellement faire varier le niveau de difficulté de la tâche, en jouant soit sur le niveau hiérarchique, soit sur la complexité des données. Cette variation du niveau de difficulté peut, lorsque pertinent, aider à préciser les conditions de cognition ou de mémoire dans lesquelles le paramètre devient suffisamment déterminant pour l'interaction. En l'occurrence, deux tâches suffisent pour faire varier le niveau de tâche. La première tâche est une tâche de bas niveau (\pex{} une tâche manipulatoire, dont la procédure est majoritairement prédéterminée et qui ne nécessite ainsi que peu de cognition, ou une tâche simple extraite d'une tâche applicative, \cad{} une tâche élémentaire). La deuxième tâche est une tâche de plus haut niveau (\cad{} nécessitant au moins un peu de cognition pour déterminer la procédure d'exécution), comme une tâche composée représentative d'une tâche applicative (\cad{} une tâche calquée sur la tâche des personnes utilisatrices finales), ou directement la tâche applicative (lorsque possible). Prendre deux tâches sera informatif pour au moins quatre raisons~:
\begin{enumerate}
\item La première tâche de bas niveau, servant de condition de base, facilitera le travail des méta-analyses pour pondérer les observations entre plusieurs études de propriétés (inclure une condition de base est recommandé lors des études de réplication \cite{hornbaek2014once}).
\item Tester la variation d'un paramètre sur deux niveaux de tâche permettra de séparer les aspects manipulatoires et solutionnels de la tâche interactive pour mieux les distinguer et les comprendre.
\item La deuxième tâche, de plus haut niveau, optimisera l'interaction plus spécifiquement pour le cas applicatif, ce qui doit rester l'objectif premier de l'étude (comme il sera expliqué plus en détail dans la section dans \hyperref[sec-serendipity-hypotheses]{la section suivante}).
\item Enfin, les tâches de haut niveau, grâce à leurs buts et procédures plus spécifiques, permettent de mieux diversifier les conditions de réplication au sein des études de propriétés. La variété des tâches parmi les réplications participera de la qualité des futures méta-analyses pour borner les conditions dans lesquelles les propriétés interactionnelles s'appliquent ou s'estompent, augmentant ainsi la confiance dans leur validité \cite{greenberg1992weak,hornbaek2014once}.
\end{enumerate}

\subsection{Faciliter la collecte de résultats disséminés}

Comme les études de propriétés devront pouvoir être répliquées par des études ultérieures, soumises à des méta-analyses ou vérifiées par des expérimentations théoriques, leur publication doit, à ces fins, s'accompagner des informations et des méta-informations adéquates~:

\begin{enumerate}
\item Les mots-clefs doivent faciliter le ciblage~: l'article doit inclure, a minima, le mot-clef \g{étude de propriété}. D'autres mots-clefs peuvent raffiner la description des études en apportant des précisions sur les prismes utilisés (\cad{} \g{prisme en entrée} ou \g{prisme en sortie}) et les paramètres variants (\pex{} \g{paramètre de vitesse}, \g{paramètre de forme} ou encore \g{paramètre de multiplexage}), ainsi que les noms de théories sous-jacentes si elles préexistent.
\item Les théories et études relatives qui préexistent doivent aussi être nommées et référencées dans l'article \cite{hornbaek2014once}, en expliquant ce qui diffère et ce qui est invariant \cite{hornbaek2014once}.
\item Pondérer les observations entre plusieurs études décrites dans plusieurs articles nécessitera de pouvoir prendre en considération leur spécificités en termes de conditions, de prototypes et de tâches, qui devront ainsi être chacun bien décrits et documentés. Une vidéo de démonstration devrait également rendre compte du prototype applicatif \cite{greenberg1992weak}, en présentant les tâches et toutes les fonctionnalités interactives impliquées.
\item En plus de décrire le dispositif expérimental et de rendre compte des résultats minutieusement, les études de propriétés nécessiteront de collecter suffisamment de données et de les analyser correctement \cite{hornbaek2014once,shepperd2018replication}.
\end{enumerate}

Avant de conclure, la section suivante illustre l'approche au travers de cinq cas applicatifs pouvant servir de source d'inspiration pour le choix de paramètres, de valeurs et de tâches, en résumant les besoins et tâches des personnes utilisatrices, les prototypes utilisés et les résultats obtenus.

\section{Illustration par cinq cas applicatifs}\label{sec-examples}

L'objectif de cette section est de rendre plus concrète l'approche proposée, de sorte à pouvoir guider ou inspirer de futures recherches. De plus, cet article ayant évité de produire des recommandations approfondies sur le choix de paramètres et de valeurs, connaître les choix faits et les résultats obtenus sur des cas applicatifs antérieurs peut servir de point d'amorce à de futures recherches. Ainsi, la présente section revient sur des études de propriétés conduites au travers de cinq cas applicatifs publiés dans la littérature (voir le récapitulatif du \autoref{tab:summary}). Ces cinq cas ont été choisis car ils ont permis d'ébaucher, développer, puis étendre l'approche proposée, ainsi que d'en vérifier la faisabilité, la viabilité, et la possible opérationnalisation dans des recherches indépendantes. Ces cinq cas permettent en même temps d'illustrer une diversité dans les champs d'application. Toutefois, d'autres études de propriétés peuvent se retrouver dans la littérature récente, sans forcément avoir considéré un cas applicatif particulier (\pex{} échelle des visualisations physiques \cite{lopez2021scaling}, pression tactile pour la notification \cite{fan2024smartwatchFR,fan2024smartwatchEN} et seuillages de vitesse pour l'intégrité physique perçue \cite{daniel2025shypins}), ou encore en ayant recouru à une quasi-expérimentation (\pex{} attrait de modalités ambiantes dans le temps \cite{daniel2021exploring}). Les résultats produits par toutes ces premières études de propriétés ouvrent la voie pour de futures réplications, puis méta-analyses et éventuelles généralisations ultérieures. La vocation de cette section est donc de guider le lecteur intéressé en décrivant cinq exemples concrets et en illustrant le formalisme des propriétés interactionnelles par des fiches synthétiques (mais en ayant omis le nom de boucle, car ce sujet mérite au préalable un approfondissement par des recherches futures).
Sinon, les prochaines sections sont une discussion des implications de l'approche proposée et la conclusion.

\renewcommand{\arraystretch}{1.2}

\begin{table*}
{\centering\small
  \begin{tabular}{@{\hspace{4pt}}r@{\hspace{4pt}}p{1.3cm}p{1.9cm}ll@{\hspace{2pt}}p{0cm}lp{1.3cm}p{1.8cm}p{0.7cm}lllp{1.5cm}l}
    
\cmidrule{1-14}

\multicolumn{5}{l}{\textbf{\textit{Cas applicatif}}} && \multicolumn{8}{l}{\textit{\textbf{Étude de propriété}}} \tabularnewline

\cmidrule{1-5}\cmidrule{7-14}

\textit{\#} & \textit{Application} & \textit{Prototype} & \textit{Année} & \textit{Réf.} && \textit{Prisme} & \textit{Paramètre} & \textit{Valeurs} & \textit{Tâches} & \textit{Année} & \textit{Réf.} & \textit{Hypothèse} & \textit{Réplication} \tabularnewline

\cmidrule{1-14}

1. &
Géosciences &
Tangible sur\newline table (GeoTUI) &
2008 &
\cite{couture2008geotui} &&
Entrée &
Multiplexage \& Forme &
Espace, Temps \& Générique,\newline Spécifique &
TBN /\newline THN\newline(TRU) & 2008 &
\cite{couture2008geotui,riviere2009phd} &
\cite[Ch. 6.1]{fitzmaurice1996phdthesis} &
\raR{}Étend \cite[Ch. 6.1]{fitzmaurice1996phdthesis}\tabularnewline[2pt]

2. &
Archéologie &
Tangible avec\newline reflètement 3D\newline (ArcheoTUI) & 2007 &
\cite{reuter2007archeotui} &&
Entrée &
Activation &
Pédalier,\newline Boutons &
TDC & 2010 &
\cite{reuter2010archeotui,riviere2010activation} &
--- &
---\tabularnewline[12pt]

3. &
Énergie\newline renouvelable &
Tangible ambiant\newline (CairnFORM) & 2019 &
\cite{daniel2019cairnform} &&
Sortie &
Vitesse &
Constante, Exponentielle, Logarithmique &
TBN /\newline THN & 2019 &
\cite{daniel2019cairnform} &
\cite{traschutz2012speed} &
Confirme~\cite{traschutz2012speed}\tabularnewline[22pt]

4. &
Chirurgie &
Lunettes transparentes\newline de réalité mixte & 2019 &
\cite{bailly2019head} && Sortie & Physicalité & Virtuel, Mixte,\newline Physique & TRU (TDA) & 2021 & \cite{bailly2021exploration,bailly2020phdthesis} & \cite{jansen2013evaluation} & Étend \cite{bailly2019head}\tabularnewline[12pt] 

5. &
Travaux publics &
Casque transparent de réalité\newline augmentée & 2021 & \cite{becher2021projectionEN} && Sortie & Indices & Projection, Projection+Grille, Grille, Aucun & TRU /\newline TBN & 2021 & \cite{becher2021projectionEN} & --- & Complète~\cite{rosales2019distance} \tabularnewline

\cmidrule{1-14}

\end{tabular}\\
    \scriptsize{\textit{Note. `Réf.' = Références. Tâches: `THN' = Tâche de Haut Niveau, `TBN' = Tâche de Bas Niveau, `TRU' = Tâche Représentative de la tâche des personnes Utilisatrices finales,\linebreak `TUF' = Tâche des personnes Utilisatrices Finales, `TDC' = Tâche à Difficulté Croissante, `TDA' = Tâche à Difficulté Alternante.}}
}
\caption{Synthèse des cinq cas applicatifs pris en illustration, par ordre d'année de l'étude de propriété.}~\label{tab:summary}
\vspace{-9pt}
\end{table*}

\subsection{Exploration du sous-sol en géosciences}

\begin{figure}[h]
  \centering
  \begin{subfigure}[b]{\linescale\linewidth}
    \centering
\def\svgwidth{8.47cm}
    {\small \begingroup \makeatletter \providecommand\color[2][]{\errmessage{(Inkscape) Color is used for the text in Inkscape, but the package 'color.sty' is not loaded}\renewcommand\color[2][]{}}\providecommand\transparent[1]{\errmessage{(Inkscape) Transparency is used (non-zero) for the text in Inkscape, but the package 'transparent.sty' is not loaded}\renewcommand\transparent[1]{}}\providecommand\rotatebox[2]{#2}\newcommand*\fsize{\dimexpr\f@size pt\relax}\newcommand*\lineheight[1]{\fontsize{\fsize}{#1\fsize}\selectfont}\ifx\svgwidth\undefined \setlength{\unitlength}{419.59192787bp}\ifx\svgscale\undefined \relax \else \setlength{\unitlength}{\unitlength * \real{\svgscale}}\fi \else \setlength{\unitlength}{\svgwidth}\fi \global\let\svgwidth\undefined \global\let\svgscale\undefined \makeatother \begin{picture}(1,0.40476247)\lineheight{1}\setlength\tabcolsep{0pt}\put(0,0){\includegraphics[width=\unitlength,page=1]{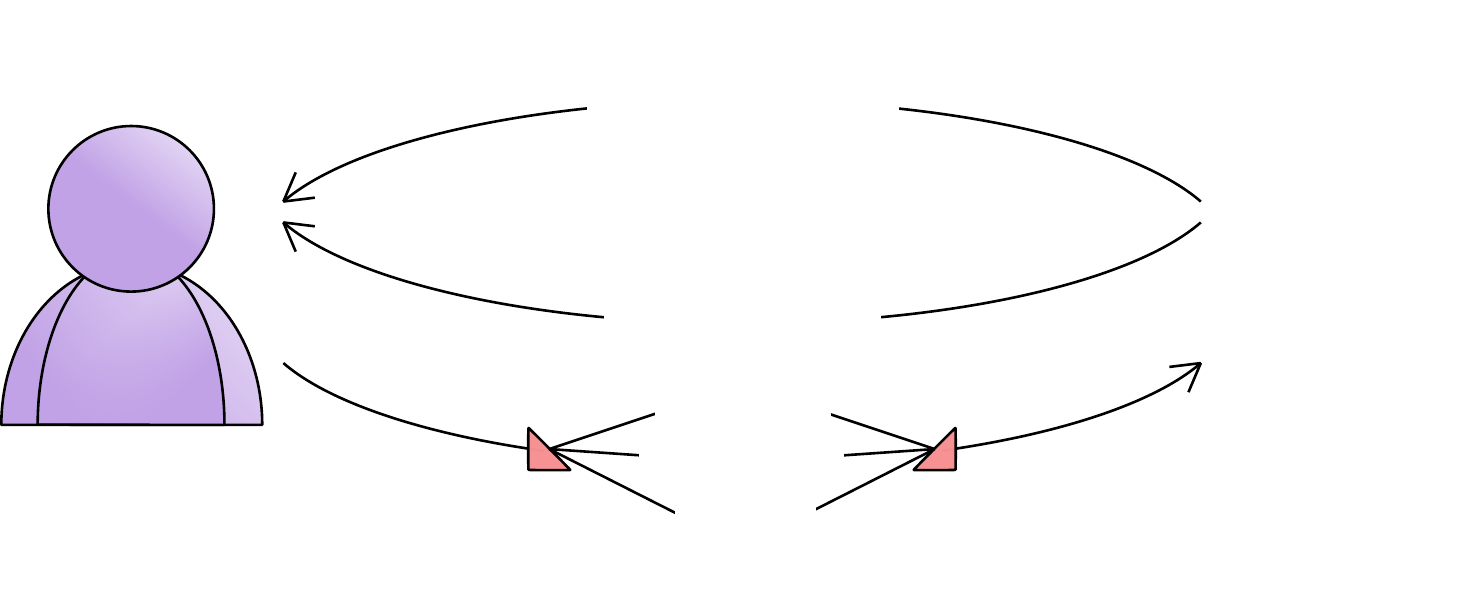}}\put(0.50746812,0.04171733){\color[rgb]{0,0,0}\makebox(0,0)[t]{\lineheight{1.25}\smash{\begin{tabular}[t]{c}\textit{Règle}\end{tabular}}}}\put(0.50971794,0.00047246){\color[rgb]{0,0,0}\makebox(0,0)[t]{\lineheight{1.25}\smash{\begin{tabular}[t]{c}\textbf{Entrée diffractée de l'outil}\end{tabular}}}}\put(0.50798292,0.34200934){\color[rgb]{0,0,0}\makebox(0,0)[t]{\lineheight{1.25}\smash{\begin{tabular}[t]{c}\textit{Croix, Ligne,}\end{tabular}}}}\put(0.50794478,0.30983524){\color[rgb]{0,0,0}\makebox(0,0)[t]{\lineheight{1.25}\smash{\begin{tabular}[t]{c}\textit{Ancres}\end{tabular}}}}\put(0.50786849,0.17756374){\color[rgb]{0,0,0}\makebox(0,0)[t]{\lineheight{1.25}\smash{\begin{tabular}[t]{c}\textit{Carte, Coupes}\end{tabular}}}}\put(0.50971792,0.21496667){\color[rgb]{0,0,0}\makebox(0,0)[t]{\lineheight{1.25}\smash{\begin{tabular}[t]{c}\textbf{Sortie des données}\end{tabular}}}}\put(0.50967978,0.37941232){\color[rgb]{0,0,0}\makebox(0,0)[t]{\lineheight{1.25}\smash{\begin{tabular}[t]{c}\textbf{Sortie de l'outil}\end{tabular}}}}\put(0.50693424,0.11321544){\color[rgb]{0,0,0}\makebox(0,0)[t]{\lineheight{1.25}\smash{\begin{tabular}[t]{c}\textit{1 palet}\end{tabular}}}}\put(0.5081545,0.07746639){\color[rgb]{0,0,0}\makebox(0,0)[t]{\lineheight{1.25}\smash{\begin{tabular}[t]{c}\textit{2 palets}\end{tabular}}}}\put(0,0){\includegraphics[width=\unitlength,page=2]{geoscience_tui-fr.pdf}}\put(0.91951814,0.10939741){\color[rgb]{0,0,0}\makebox(0,0)[t]{\smash{\begin{tabular}[t]{c}Tangible\\sur table\\de vision-\\projection\end{tabular}}}}\put(0.08914265,0.06482075){\color[rgb]{0,0,0}\makebox(0,0)[t]{\smash{\begin{tabular}[t]{c}Personne\\utilisatrice\end{tabular}}}}\end{picture}\endgroup  }
    \caption{Conditions d'interaction tangible sur table.}
    \label{fig:loops-prisms-geotui-tui}
    \Description{A user is on the left. A vision-projection tabletop is on the right. An arrow from the user to the tabletop is diffracted between two prisms, thus going through three input tools: a one-puck prop, a two-puck prop, and a ruler prop. Likewise, two arrows go from the tabletop to the user. The first arrow is tool output: cross, line, and anchors. The second arrow is datum output: map and slice.}
  \end{subfigure}
  \begin{subfigure}[b]{\linescale\linewidth}
    \centering
    \vspace{12pt}
\def\svgwidth{8.47cm}
    {\small \begingroup \makeatletter \providecommand\color[2][]{\errmessage{(Inkscape) Color is used for the text in Inkscape, but the package 'color.sty' is not loaded}\renewcommand\color[2][]{}}\providecommand\transparent[1]{\errmessage{(Inkscape) Transparency is used (non-zero) for the text in Inkscape, but the package 'transparent.sty' is not loaded}\renewcommand\transparent[1]{}}\providecommand\rotatebox[2]{#2}\newcommand*\fsize{\dimexpr\f@size pt\relax}\newcommand*\lineheight[1]{\fontsize{\fsize}{#1\fsize}\selectfont}\ifx\svgwidth\undefined \setlength{\unitlength}{422.46010072bp}\ifx\svgscale\undefined \relax \else \setlength{\unitlength}{\unitlength * \real{\svgscale}}\fi \else \setlength{\unitlength}{\svgwidth}\fi \global\let\svgwidth\undefined \global\let\svgscale\undefined \makeatother \begin{picture}(1,0.37691694)\lineheight{1}\setlength\tabcolsep{0pt}\put(0,0){\includegraphics[width=\unitlength,page=1]{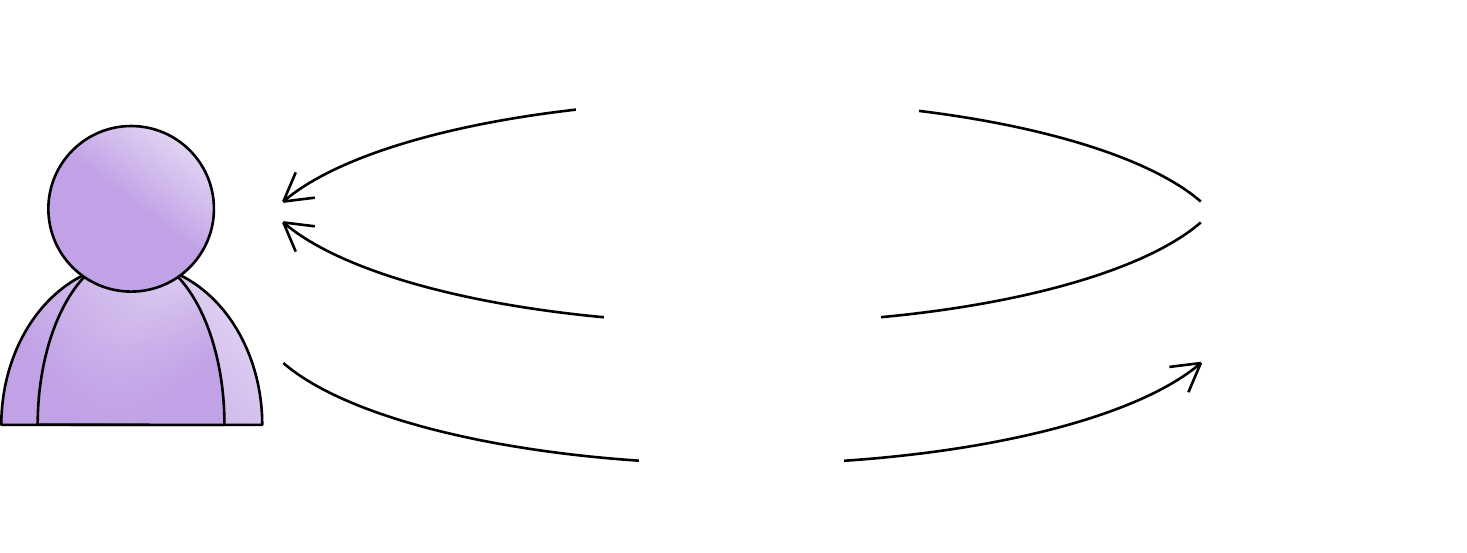}}\put(0.50625735,0.01442869){\color[rgb]{0,0,0}\makebox(0,0)[t]{\lineheight{1.25}\smash{\begin{tabular}[t]{c}\textbf{Entrée de l'outil}\end{tabular}}}}\put(0.50453411,0.31458985){\color[rgb]{0,0,0}\makebox(0,0)[t]{\lineheight{1.25}\smash{\begin{tabular}[t]{c}\textit{Curseur, Ligne,}\end{tabular}}}}\put(0.50449623,0.28263419){\color[rgb]{0,0,0}\makebox(0,0)[t]{\lineheight{1.25}\smash{\begin{tabular}[t]{c}\textit{Ancres}\end{tabular}}}}\put(0.50442047,0.15126071){\color[rgb]{0,0,0}\makebox(0,0)[t]{\lineheight{1.25}\smash{\begin{tabular}[t]{c}\textit{Carte, Coupes}\end{tabular}}}}\put(0.50625734,0.18840971){\color[rgb]{0,0,0}\makebox(0,0)[t]{\lineheight{1.25}\smash{\begin{tabular}[t]{c}\textbf{Sortie de données}\end{tabular}}}}\put(0.50621946,0.35173889){\color[rgb]{0,0,0}\makebox(0,0)[t]{\lineheight{1.25}\smash{\begin{tabular}[t]{c}\textbf{Sortie de l'outil}\end{tabular}}}}\put(0.50470453,0.05184294){\color[rgb]{0,0,0}\makebox(0,0)[t]{\lineheight{1.25}\smash{\begin{tabular}[t]{c}\textit{Souris}\end{tabular}}}}\put(0,0){\includegraphics[width=\unitlength,page=2]{geoscience_gui-fr.pdf}}\put(0.91744319,0.05981291){\color[rgb]{0,0,0}\makebox(0,0)[t]{\lineheight{1.10000002}\smash{\begin{tabular}[t]{c}Interface\\graphique\end{tabular}}}}\put(0.08853744,0.03928316){\color[rgb]{0,0,0}\makebox(0,0)[t]{\lineheight{1.10000002}\smash{\begin{tabular}[t]{c}Personne\\utilisatrice\end{tabular}}}}\put(0,0){\includegraphics[width=\unitlength,page=3]{geoscience_gui-fr.pdf}}\end{picture}\endgroup  }
    \caption{Conditions d'interaction graphique.}
    \label{fig:loops-prisms-geotui-gui}
    \Description{A user is on the left. A desktop GUI is on the right. An arrow from the user to the desktop is diffracted between two prisms, thus going through three input tools: a one-puck prop, a two-puck prop, and a ruler prop. Likewise, two output arrows go from the desktop to the user. The first arrow is tools output: cross, line, and anchors. The second arrow is data output: map and slice.}
  \end{subfigure}
  \caption{Boucles testées pour la navigation depuis une carte \cite{couture2008geotui}. Cette première étude de propriété s'entremêlait avec une comparaison de technologies, de sorte à avoir l'interaction graphique comme condition de référence. Deux niveaux de tâches ont été testés au travers de ces boucles.}
  \label{fig:loops-prisms-geotui}
\end{figure}

En 2008, le prototype GeoTUI a permis d'explorer une tâche de navigation sur une carte 2D \cite{couture2008geotui,riviere2009phd}.
Pour répondre aux besoins des géophysiciens de collaboration en coprésence, lorsqu'ils font des hypothèses sur la composition du sous-sol, cette table interactive tangible affiche une carte de terrain en vue de dessus. La sélection d'une ligne depuis cette carte permet de trancher dans le modèle de sous-sol et obtenir un plan de coupe. Lors d'une deuxième étude expérimentale \citetext{\citealp{couture2008geotui}, \citealp[Ch.~3]{riviere2009phd}}, un prisme interactionnel en entrée a servi à diffracter l'interaction pour la sélection d'une ligne de coupe selon le multiplexage et la forme des interacteurs physiques utilisés (voir \autoref{fig:loops-prisms-geotui-tui}). Cette étude a permis de préciser les conditions d'apparition d'une hypothèse antérieure sur la manipulation d'interacteurs tangibles \cite[Ch.~6.1]{fitzmaurice1996phdthesis}, comme synthétisé par la fiche de propriété du \autoref{tab:propGEO}.

\ficheGEO{h}{Propriété observée sur un cas en géosciences. Étudier une tâche applicative a permis d'étendre une hypothèse qui était restée insensible à des tâches de laboratoire uniquement manipulatoires.}

\subsection{Assemblages de fragments en archéologie}

\begin{figure}[h]
  \centering
\def\svgwidth{8.47cm}
  {\small \begingroup \makeatletter \providecommand\color[2][]{\errmessage{(Inkscape) Color is used for the text in Inkscape, but the package 'color.sty' is not loaded}\renewcommand\color[2][]{}}\providecommand\transparent[1]{\errmessage{(Inkscape) Transparency is used (non-zero) for the text in Inkscape, but the package 'transparent.sty' is not loaded}\renewcommand\transparent[1]{}}\providecommand\rotatebox[2]{#2}\newcommand*\fsize{\dimexpr\f@size pt\relax}\newcommand*\lineheight[1]{\fontsize{\fsize}{#1\fsize}\selectfont}\ifx\svgwidth\undefined \setlength{\unitlength}{412.75501359bp}\ifx\svgscale\undefined \relax \else \setlength{\unitlength}{\unitlength * \real{\svgscale}}\fi \else \setlength{\unitlength}{\svgwidth}\fi \global\let\svgwidth\undefined \global\let\svgscale\undefined \makeatother \begin{picture}(1,0.42087644)\lineheight{1}\setlength\tabcolsep{0pt}\put(0,0){\includegraphics[width=\unitlength,page=1]{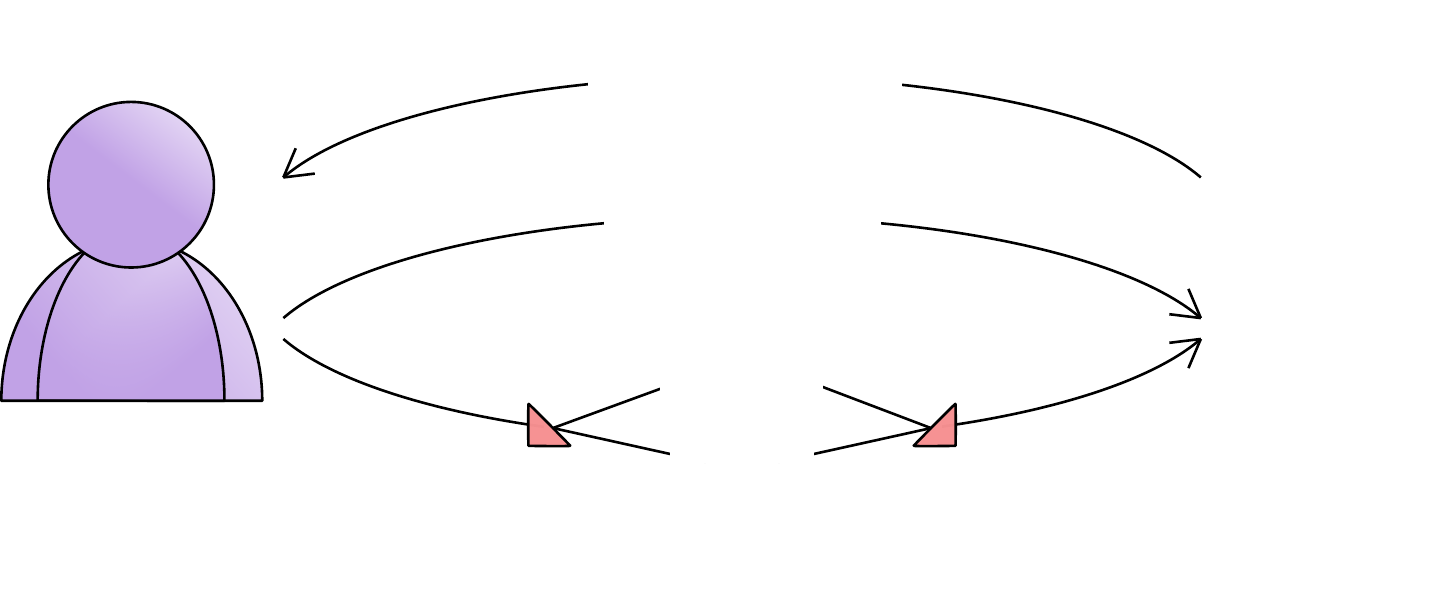}}\put(0.51816095,0.04123872){\color[rgb]{0,0,0}\makebox(0,0)[t]{\smash{\begin{tabular}[t]{c}\textbf{Activation diffractée en entrée}\\\textbf{(mécanisme d'embrayage)}\end{tabular}}}}\put(0.5163972,0.3520917){\color[rgb]{0,0,0}\makebox(0,0)[t]{\lineheight{1.25}\smash{\begin{tabular}[t]{c}\textit{Fragments 3D}\end{tabular}}}}\put(0.51628088,0.25760457){\color[rgb]{0,0,0}\makebox(0,0)[t]{\lineheight{1.25}\smash{\begin{tabular}[t]{c}\textit{Deux}\end{tabular}}}}\put(0.51816094,0.29199293){\color[rgb]{0,0,0}\makebox(0,0)[t]{\lineheight{1.25}\smash{\begin{tabular}[t]{c}\textbf{Entrée de données}\end{tabular}}}}\put(0.51812217,0.39374832){\color[rgb]{0,0,0}\makebox(0,0)[t]{\lineheight{1.25}\smash{\begin{tabular}[t]{c}\textbf{Sortie de données}\end{tabular}}}}\put(0,0){\includegraphics[width=\unitlength,page=2]{archaeology-fr.pdf}}\put(0.51657162,0.14131271){\color[rgb]{0,0,0}\makebox(0,0)[t]{\lineheight{1.25}\smash{\begin{tabular}[t]{c}\textit{Pédalier}\end{tabular}}}}\put(0.51657162,0.08680091){\color[rgb]{0,0,0}\makebox(0,0)[t]{\lineheight{1.25}\smash{\begin{tabular}[t]{c}\textit{Boutons}\end{tabular}}}}\put(0.93901498,0.11312888){\color[rgb]{0,0,0}\makebox(0,0)[t]{\smash{\begin{tabular}[t]{c}Bureau\\3D\end{tabular}}}}\put(0.51628088,0.22853161){\color[rgb]{0,0,0}\makebox(0,0)[t]{\lineheight{1.25}\smash{\begin{tabular}[t]{c}\textit{interacteurs 6DOF}\end{tabular}}}}\put(0.09061922,0.09211641){\color[rgb]{0,0,0}\makebox(0,0)[t]{\smash{\begin{tabular}[t]{c}Personne\\utilisatrice\end{tabular}}}}\end{picture}\endgroup  }
  \caption{Boucle testée pour l'assemblage virtuel \cite{reuter2010archeotui,riviere2010activation}. La particularité de cette étude est d'avoir diffracté l'activation d'une modalité de contrôle. De plus, la difficulté de la tâche allait croissant de par le jeu de données.}
  \Description{A user is on the left. A desktop 3DUI is on the right. Two input arrows go from the user to the desktop. The first arrow is datum input: two props. The second input arrow is diffracted between two prisms, thus going through two input activation mechanisms (\ie{} clutching): foot pedals and hand buttons. Finally, an arrow from the desktop to the user is datum output: 3D fragments.}
  \label{fig:loops-prisms-archeotui}
\end{figure}

En 2010, une étude de propriété avec le prototype ArcheoTUI a exploré une technique d'assemblage virtuel \cite{reuter2010archeotui,riviere2010activation}. Les assemblages virtuels permettent aux archéologues de tester des hypothèses sur des fragments qui peuvent être trop friables, trop énormes, ou bien hors de leur possession. Cependant, de tels assemblages sont irréalisables avec les dispositifs d'interaction traditionnels. ArcheoTUI rend ces assemblages possibles par transposition à deux fragments archéologiques 3D, affichés sur un écran, de la position et de l'orientation de deux interacteurs, manipulés dans l'espace, une fois que deux pédales sont embrayées par action des pieds \cite{reuter2007archeotui}. Lors d'une nouvelle étude expérimentale, un prisme interactionnel a été placé en entrée pour diffracter l'interaction au niveau du dispositif de débrayage (voir \autoref{fig:loops-prisms-archeotui}). Les observations montrent une transformation totale du geste d'assemblage en fonction de l'activation de modalité, comme résumé par la fiche de propriété du \autoref{tab:propARCHEO}.

\ficheARCHEO{h}{Propriété observée sur un cas en archéologie. Avoir modifié l'activation de modalité a complètement changé la nature des mouvements d'assemblages, devenus brefs et axiaux, sans autre explication que le blocage des poignets entraînant des mouvements des avant-bras plutôt que des mains et par les doigts.}

\subsection{Maximisation des énergies renouvelables dans des espaces de bureaux}

\begin{figure}[h]
  \centering
\def\svgwidth{8.47cm}
  {\small \begingroup \makeatletter \providecommand\color[2][]{\errmessage{(Inkscape) Color is used for the text in Inkscape, but the package 'color.sty' is not loaded}\renewcommand\color[2][]{}}\providecommand\transparent[1]{\errmessage{(Inkscape) Transparency is used (non-zero) for the text in Inkscape, but the package 'transparent.sty' is not loaded}\renewcommand\transparent[1]{}}\providecommand\rotatebox[2]{#2}\newcommand*\fsize{\dimexpr\f@size pt\relax}\newcommand*\lineheight[1]{\fontsize{\fsize}{#1\fsize}\selectfont}\ifx\svgwidth\undefined \setlength{\unitlength}{428.84651562bp}\ifx\svgscale\undefined \relax \else \setlength{\unitlength}{\unitlength * \real{\svgscale}}\fi \else \setlength{\unitlength}{\svgwidth}\fi \global\let\svgwidth\undefined \global\let\svgscale\undefined \makeatother \begin{picture}(1,0.40100636)\lineheight{1}\setlength\tabcolsep{0pt}\put(0,0){\includegraphics[width=\unitlength,page=1]{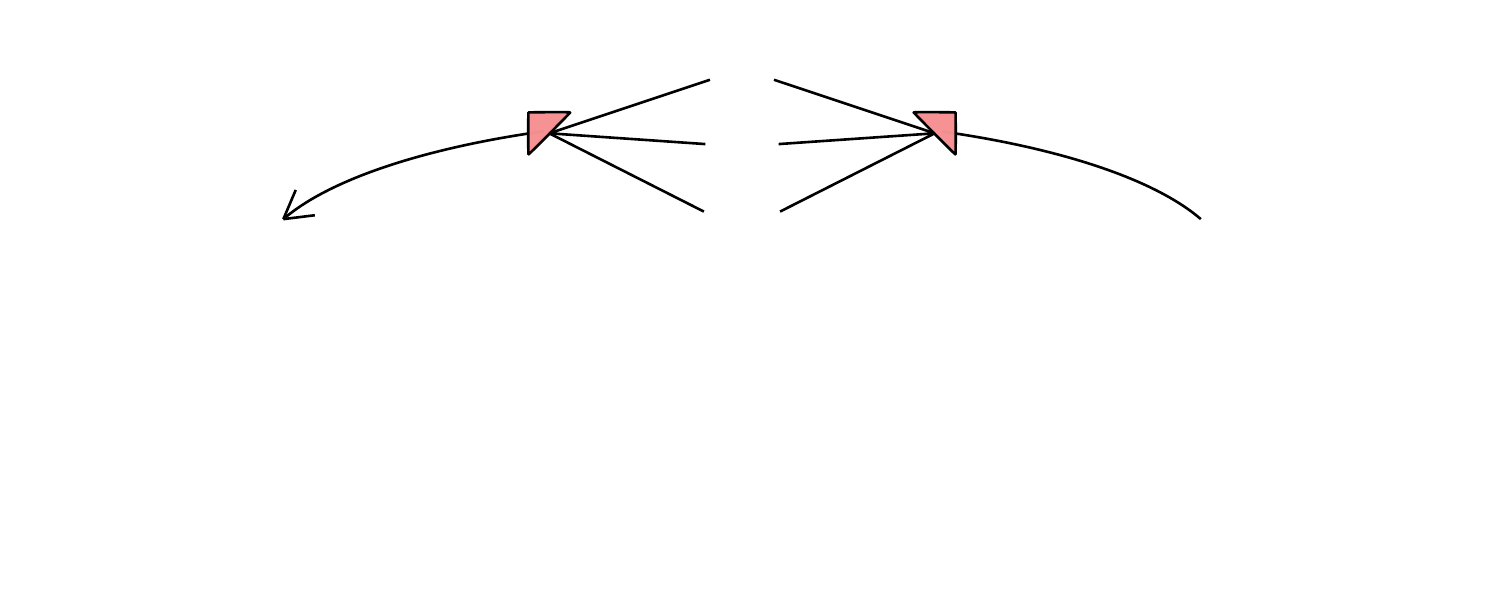}}\put(0.49868079,0.37489616){\color[rgb]{0,0,0}\makebox(0,0)[t]{\lineheight{1.25}\smash{\begin{tabular}[t]{c}\textbf{Sortie diffractée des données (Profil de vitesse)}\end{tabular}}}}\put(0,0){\includegraphics[width=\unitlength,page=2]{energy-fr.pdf}}\put(0.49847645,0.29423925){\color[rgb]{0,0,0}\makebox(0,0)[t]{\lineheight{1.25}\smash{\begin{tabular}[t]{c}\textit{Exp.}\end{tabular}}}}\put(0,0){\includegraphics[width=\unitlength,page=3]{energy-fr.pdf}}\put(0.88324988,0.10472772){\makebox(0,0)[t]{\smash{\begin{tabular}[t]{c}Histogramme\\physique\\à anneaux\\dynamiques\end{tabular}}}}\put(0.49847645,0.33621229){\color[rgb]{0,0,0}\makebox(0,0)[t]{\lineheight{1.25}\smash{\begin{tabular}[t]{c}\textit{Const.}\end{tabular}}}}\put(0.49847645,0.25226615){\color[rgb]{0,0,0}\makebox(0,0)[t]{\lineheight{1.25}\smash{\begin{tabular}[t]{c}\textit{Log.}\end{tabular}}}}\put(0.08721893,0.05660028){\color[rgb]{0,0,0}\makebox(0,0)[t]{\smash{\begin{tabular}[t]{c}Personne\\utilisatrice\end{tabular}}}}\put(0,0){\includegraphics[width=\unitlength,page=4]{energy-fr.pdf}}\end{picture}\endgroup  }
  \caption{Boucle testée pour notifier sur un lieu de travail \cite{daniel2019cairnform}. La particularité de cette étude est d'avoir diffracté d'une communication unilatérale sollicitant la vision périphérique, qui a été testée sur deux tâches de bas niveau, de perception et de détection, simultanément à des tâches de distraction effectuées en vision centrale.}
  \Description{A user is on the left. CairnFORM prototype is on the right. An arrow from CairnFORM to the user is diffracted between two prisms, thus going through three datum outputs speed profiles: constant, exponential, and logarithmic.}
  \label{fig:loops-prisms-cairnform}
\end{figure}

En 2019, le prototype CairnFORM a été conçu pour la maximisation de la consommation d'énergie renouvelable, pour des lieux de travail, en s'appuyant sur des moyens de stockage et des prévisions d'énergie renouvelable \cite{daniel2019cairnform}. CairnFORM est un histogramme cylindrique à changement de forme, dont l'expansion de dix anneaux est utilisée pour représenter les taux de disponibilité d'énergie renouvelable sur une période de dix heures. Concevoir des notifications ambiantes sur les espaces de travail est délicat parce que les notifications doivent pouvoir être perçues -- pour la réussite du décalage de la consommation d'énergie -- mais sans perturber, ni irriter, les personnes concentrées sur leurs activités professionnelles. Dans l'objectif de respecter ces exigences, l'amélioration de la perception des personnes utilisatrices a été mesurée en faisant varier l'expansion des diamètres des anneaux selon trois profils de vitesse, déterminés avec une part d'intuition et quelques connaissances en biologie. Dans cette optique, un prisme interactionnel a été placé en sortie pour diffracter la perception des mouvements des anneaux~: le paramètre variant était la vitesse d'expansion et les trois valeurs prises étaient des profils de vitesse constante, exponentielle et logarithmique (voir \autoref{fig:loops-prisms-cairnform}). Cette étude de propriété, synthétisée par la fiche du \autoref{tab:propENERGY}, a permis de confirmer des observations précédentes sur la perception de l'accélération et de la décélération en zone périphérique, une explication de ces différences de perception pouvant résider dans la perception du flux optique pendant les déplacements \cite{traschutz2012speed}.

\ficheENERGY{h}{Propriété observée sur un cas en énergétique. Rechercher un mouvement, d'expansion et de rétraction, suffisamment perceptible mais sans perturber, pour notifier sur un lieu de travail, a permis de répéter une observation antérieure sur la perception en vision périphérique.}

\subsection{Navigation dans un plan d'action en chirurgie augmentée}

\begin{figure}[h]
  \centering
\def\svgwidth{8.47cm}
  {\small \begingroup \makeatletter \providecommand\color[2][]{\errmessage{(Inkscape) Color is used for the text in Inkscape, but the package 'color.sty' is not loaded}\renewcommand\color[2][]{}}\providecommand\transparent[1]{\errmessage{(Inkscape) Transparency is used (non-zero) for the text in Inkscape, but the package 'transparent.sty' is not loaded}\renewcommand\transparent[1]{}}\providecommand\rotatebox[2]{#2}\newcommand*\fsize{\dimexpr\f@size pt\relax}\newcommand*\lineheight[1]{\fontsize{\fsize}{#1\fsize}\selectfont}\ifx\svgwidth\undefined \setlength{\unitlength}{416.70419738bp}\ifx\svgscale\undefined \relax \else \setlength{\unitlength}{\unitlength * \real{\svgscale}}\fi \else \setlength{\unitlength}{\svgwidth}\fi \global\let\svgwidth\undefined \global\let\svgscale\undefined \makeatother \begin{picture}(1,0.43507982)\lineheight{1}\setlength\tabcolsep{0pt}\put(0,0){\includegraphics[width=\unitlength,page=1]{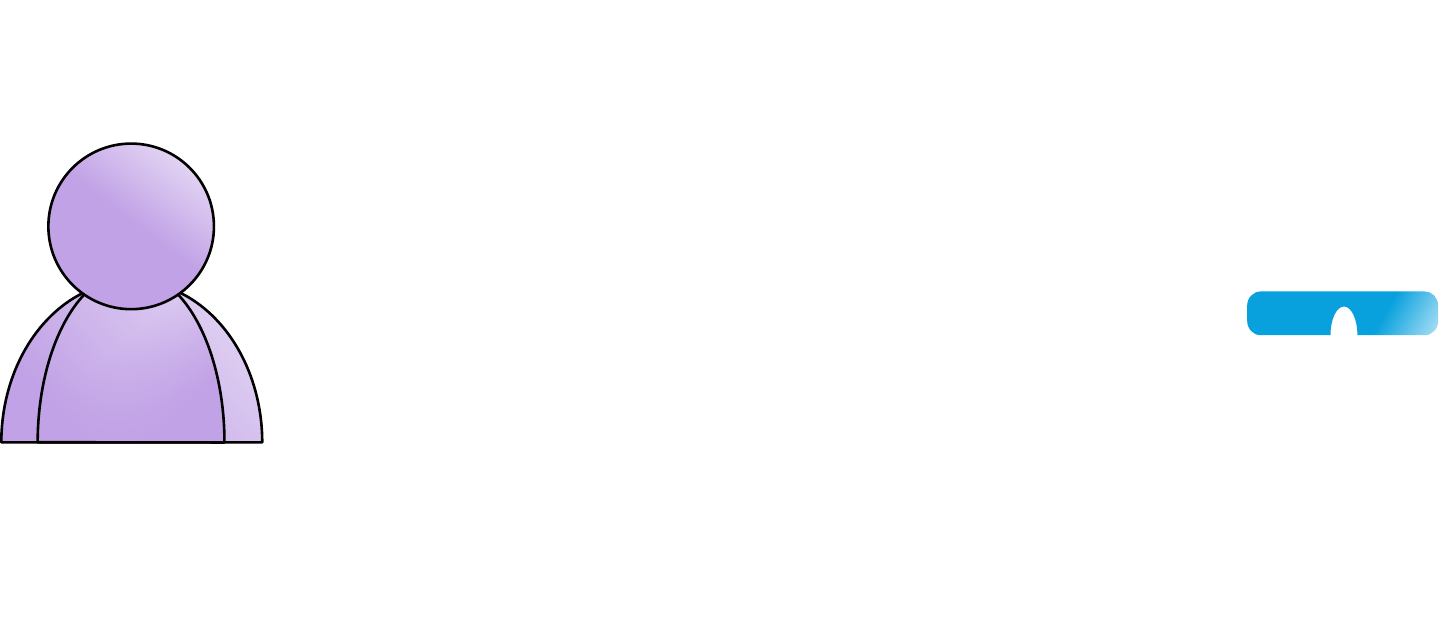}}\put(0.0897604,0.0806381){\color[rgb]{0,0,0}\makebox(0,0)[t]{\smash{\begin{tabular}[t]{c}Personne\\utilisatrice\end{tabular}}}}\put(0,0){\includegraphics[width=\unitlength,page=2]{surgery-fr.pdf}}\put(0.51321182,0.40820879){\color[rgb]{0,0,0}\makebox(0,0)[t]{\lineheight{1.25}\smash{\begin{tabular}[t]{c}\textbf{Sortie diffractée du menu}\end{tabular}}}}\put(0,0){\includegraphics[width=\unitlength,page=3]{surgery-fr.pdf}}\put(0.51300153,0.36479804){\color[rgb]{0,0,0}\makebox(0,0)[t]{\lineheight{1.25}\smash{\begin{tabular}[t]{c}\textit{Virtuel}\end{tabular}}}}\put(0,0){\includegraphics[width=\unitlength,page=4]{surgery-fr.pdf}}\put(0.51325024,0.03744194){\color[rgb]{0,0,0}\makebox(0,0)[t]{\lineheight{1.25}\smash{\begin{tabular}[t]{c}\textbf{Entrée du pointeur}\end{tabular}}}}\put(0.51138798,0.19056271){\color[rgb]{0,0,0}\makebox(0,0)[t]{\lineheight{1.25}\smash{\begin{tabular}[t]{c}\textit{Curseur}\end{tabular}}}}\put(0.51325023,0.22462517){\color[rgb]{0,0,0}\makebox(0,0)[t]{\lineheight{1.25}\smash{\begin{tabular}[t]{c}\textbf{Sortie du pointeur}\end{tabular}}}}\put(0.51167597,0.11856913){\color[rgb]{0,0,0}\makebox(0,0)[t]{\smash{\begin{tabular}[t]{c}\textit{Position du curseur}\\\textit{(Mouvements de tête)}\end{tabular}}}}\put(0.83215768,0.10126665){\color[rgb]{0,0,0}\makebox(0,0)[lt]{\smash{\begin{tabular}[t]{l}Lunettes de\\réalité mixte\end{tabular}}}}\put(0.51300153,0.29280451){\color[rgb]{0,0,0}\makebox(0,0)[t]{\lineheight{1.25}\smash{\begin{tabular}[t]{c}\textit{Physique}\end{tabular}}}}\put(0.51300153,0.3288013){\color[rgb]{0,0,0}\makebox(0,0)[t]{\lineheight{1.25}\smash{\begin{tabular}[t]{c}\textit{Mixte}\end{tabular}}}}\put(0,0){\includegraphics[width=\unitlength,page=5]{surgery-fr.pdf}}\end{picture}\endgroup  }
  \caption{Boucle testée pour la sélection dans des menus en salles opératoires \cite{bailly2021exploration}. Dans cette étude, le pointage dans un menu était alterné avec deux tâches de distraction de plus haut niveau, consistant en un raisonnement mathématique.}
  \Description{A user is on the left. A pair of see-through mixed reality glasses is on the right. An arrow from the user to the glasses if pointer input: cursor position (head motion). Two arrows go from the glasses to the user. The first arrow is the pointer output: cursor. The second arrow is diffracted between two prisms, thus going through three menu outputs: virtual, mixed, and physical.}
  \label{fig:loops-prisms-surgery}
\end{figure}

En 2021, le recours à des lunettes de réalité augmentée a été envisagé dans le domaine chirurgical \cite{bailly2021exploration,bailly2020phdthesis}. Générer des entrées système par mouvements de la tête répond aux contraintes des chirurgiens de respecter un milieu stérile, de conserver leurs mains libres et d'évoluer dans un environnement bruyant et déjà encombré. Ainsi, un curseur contrôlé par mouvements de tête est utilisé comme pointeur pour la sélection dans des menus. Une étude de propriété a testé trois variantes pour la physicalité d'un tel menu en plaçant un prisme en sortie (voir \autoref{fig:loops-prisms-surgery}) et a permis d'obtenir des résultats sur la physicalité des cibles pour le pointage, comme synthétisé par la fiche de propriété du \autoref{tab:propSURGERY}.

\ficheSURGERY{h}{Propriété observée sur un cas en chirurgie. Une étude, pour répondre aux contraintes de milieu stérile de ce cas, a permis d'étendre à des menus, des résultats antérieurement observés sur des histogrammes.}

\subsection{Vision souterraine pour les travaux publics}

\begin{figure}[h]
  \centering
\def\svgwidth{8.47cm}
  {\small \begingroup \makeatletter \providecommand\color[2][]{\errmessage{(Inkscape) Color is used for the text in Inkscape, but the package 'color.sty' is not loaded}\renewcommand\color[2][]{}}\providecommand\transparent[1]{\errmessage{(Inkscape) Transparency is used (non-zero) for the text in Inkscape, but the package 'transparent.sty' is not loaded}\renewcommand\transparent[1]{}}\providecommand\rotatebox[2]{#2}\newcommand*\fsize{\dimexpr\f@size pt\relax}\newcommand*\lineheight[1]{\fontsize{\fsize}{#1\fsize}\selectfont}\ifx\svgwidth\undefined \setlength{\unitlength}{422.86196046bp}\ifx\svgscale\undefined \relax \else \setlength{\unitlength}{\unitlength * \real{\svgscale}}\fi \else \setlength{\unitlength}{\svgwidth}\fi \global\let\svgwidth\undefined \global\let\svgscale\undefined \makeatother \begin{picture}(1,0.38285683)\lineheight{1}\setlength\tabcolsep{0pt}\put(0.0884533,0.03357654){\color[rgb]{0,0,0}\makebox(0,0)[t]{\smash{\begin{tabular}[t]{c}Personne\\utilisatrice\end{tabular}}}}\put(0,0){\includegraphics[width=\unitlength,page=1]{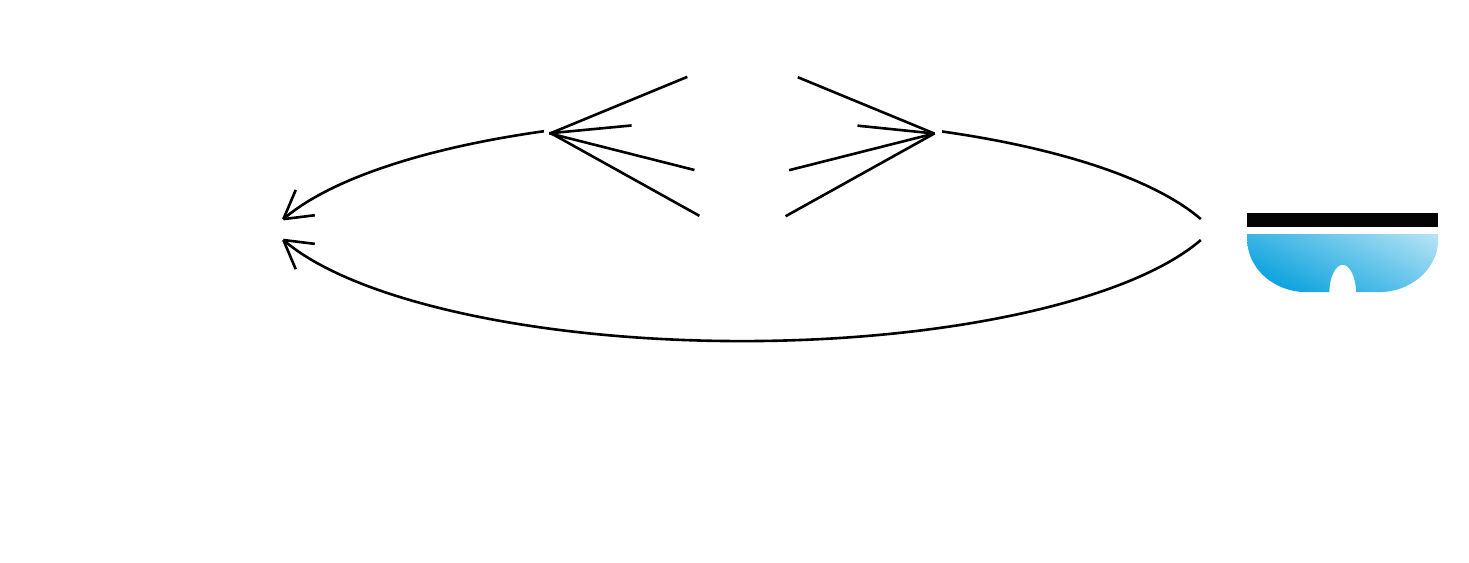}}\put(0.50573838,0.3563771){\color[rgb]{0,0,0}\makebox(0,0)[t]{\lineheight{1.25}\smash{\begin{tabular}[t]{c}\textbf{Sortie diffractée des indices}\end{tabular}}}}\put(0,0){\includegraphics[width=\unitlength,page=2]{public_works-fr.pdf}}\put(0.50553115,0.26038965){\color[rgb]{0,0,0}\makebox(0,0)[t]{\smash{\begin{tabular}[t]{c}\textit{Grille}\end{tabular}}}}\put(0,0){\includegraphics[width=\unitlength,page=3]{public_works-fr.pdf}}\put(0.5039411,0.14544767){\color[rgb]{0,0,0}\makebox(0,0)[t]{\lineheight{1.25}\smash{\begin{tabular}[t]{c}\textit{Cubes 3D}\end{tabular}}}}\put(0.50577623,0.1080689){\color[rgb]{0,0,0}\makebox(0,0)[t]{\lineheight{1.25}\smash{\begin{tabular}[t]{c}\textbf{Sortie de données}\end{tabular}}}}\put(0,0){\includegraphics[width=\unitlength,page=4]{public_works-fr.pdf}}\put(0.90695855,0.12475328){\color[rgb]{0,0,0}\makebox(0,0)[t]{\smash{\begin{tabular}[t]{c}Casque\\de réalité\\augmentée\end{tabular}}}}\put(0.50553115,0.29231498){\color[rgb]{0,0,0}\makebox(0,0)[t]{\smash{\begin{tabular}[t]{c}\textit{Proj.+Grille}\end{tabular}}}}\put(0.50553115,0.32424027){\color[rgb]{0,0,0}\makebox(0,0)[t]{\smash{\begin{tabular}[t]{c}\textit{Proj.}\end{tabular}}}}\put(0.50553115,0.23201157){\color[rgb]{0,0,0}\makebox(0,0)[t]{\smash{\begin{tabular}[t]{c}\textit{Aucun}\end{tabular}}}}\put(0,0){\includegraphics[width=\unitlength,page=5]{public_works-fr.pdf}}\end{picture}\endgroup  }
  \caption{Boucle testée pour la vision souterraine \cite{becher2021projectionFR,becher2021projectionEN}. Trois tâches ont été testées au travers de cette boucle, dont une représentative de la tâche applicative consistant à effectuer une action en-dehors du système (marcher pour déposer une croix au sol).} \Description{A user is on the left. A see-through augmented reality headset is on the right. Two arrows go from the headset to the user. The first arrow is data output: 3D cubes. The second arrow is diffracted between two prisms, thus going through four cue outputs: projection, projection + grid, grid, and none.}
  \label{fig:loops-prisms-underground}
\end{figure}

Toujours en 2021, l'opportunité d'utiliser un casque de réalité augmentée a été explorée, au travers d'une tâche de marquage, pour les cartes souterraines utilisées pour les travaux publics \cite{becher2021projectionFR,becher2021projectionEN}. Les travaux publics s'efforcent de prévenir les incidents liés aux installations souterraines (\pex{} les réseaux d'électricité, d'eau et de gaz) en les référençant sur des cartes de profondeur. Cependant, ces cartes restent difficiles à interpréter depuis des représentations 2D (\pex{} papier ou écrans) une fois sur le terrain. Une façon d'y remédier est d'augmenter la vision des travailleurs en ajoutant, à leur perception du lieu en chantier, des informations du souterrain. Une étude de propriété a cherché à trouver les meilleurs indices à ajouter aux augmentations pour percevoir l'altitude et la distance des installations souterraines (voir \autoref{fig:loops-prisms-underground}). Les suppositions étaient fondées sur le rôle de la surface du sol dans la perception de la distance, tout comme pour une étude antérieure qui testait la perception de la profondeur sans ajout d'indices \cite{rosales2019distance}. Les observations montrent que les ombres ancrées permettent une meilleure performance, avec ou sans la grille, que la grille seule ou que sans aucun indice, comme résumé par la fiche de propriété du \autoref{tab:propPUBLICWORKS}.

\fichePUBLICWORKS{h}{Propriété observée sur un cas en travaux publics. Ce cas obtient un autre meilleur indice de perception 3D, lors d'une tâche applicative, plutôt qu'une tâche de laboratoire visant à classer des objets.}

\section{Discussion}\label{sec-discussion}

La présente section discute l'approche proposée, au regard de son potentiel à engendrer des réplications et générer des publications.

\subsection{Initier et maintenir les réplications}

Depuis plusieurs décennies, le fossé entre science et ingénierie en IHM \cite{kostakos2015hole,reeves2015locating} soulève la question du manque de réplication des résultats \cite{cockburn2020threats,greenberg1992weak,greenberg2008usability,hornbaek2014once,hornbaek2015wrong,molich2018usability} conduisant à une surgénéralisation de résultats isolés \cite{greenberg1992weak}, et qui semble continuer de pâtir d'une faible priorisation comparé au développement de nouvelles technologies et de démonstration de leur utilité \cite{greenberg1992weak,hornbaek2014once}. Cependant, cette situation persiste potentiellement, non pas par manque de volonté, mais par l'absence d'une méthodologie adaptée et adéquate. Dans cette veine, le présent article propose l'adjonction d'études expérimentales calibrées selon un nouveau format investiguant l'interaction (les études de propriétés) lors du traitement de cas applicatifs, de sorte à multiplier et diversifier les observations de propriétés interactionnelles.

Les études de propriété ont pour but d'étudier ``l'interaction plutôt que les interfaces'' \cite{mbl2004designing}, contribuant ainsi à définir ``\textit{les lois les plus appropriées à embarquer pour régir un système interactif}'' \cite{greenberg1992weak}. Pour ce faire, les boucles d'interaction sont diffractées en un unique point, en imaginant l'utilisation d'un prisme interactionnel, pour accomplir des tâches de difficultés éventuellement variables, de sorte à observer les implications interactionnelles, fournissant ainsi un ensemble d'éléments de preuve. Ces multiples éléments factuels, obtenus dans les conditions variées de multiples cas applicatifs, peuvent ensuite aider à borner les conditions d'apparition d'un résultat \cite{hornbaek2014once}, à alimenter une méta-analyse, puis à le généraliser, en le reliant à une théorie \cite{greenberg1992weak}, permettant ainsi de l'étendre à des contextes plus larges. Dans le cas où aucune théorie préexiste, les éléments factuels accumulés peuvent aussi servir à initier des approches d'induction de théories.
Ainsi, cumuler systématiquement des ensembles d'observations calibrées, émanant des études de propriétés, et les relier aux théories sous-jacentes, permettra alors aux études empiriques de pouvoir survivre aux prototypes, technologies et tâches impliquées \cite{greenberg1992weak} et de remettre en question les idées préétablies \cite{hornbaek2014once}. La présente approche est donc scientifiquement fondée.

Au-delà d'apporter ``formalisme et rigueur'' \cite{greenberg1992weak}, des études de propriétés qui tentent, dans un premier temps, d'optimiser l'interaction avec les prototypes de cas applicatifs, participent aussi au développement technologique et répondent donc à la ``course à la sortie de nouveaux produits'' \cite{greenberg1992weak}. Ainsi, la présente approche est également économiquement fondée. De plus, la nécessité de rendre les études expérimentales plus informatives, en manipulant les paramètres des interfaces pour améliorer leur performance, a déjà été exprimée \cite{hornbaek2015wrong}.

Enfin, avoir la simple finalité d'optimiser des prototypes applicatifs permet de prendre le risque de travailler sans hypothèse théorique préalable -- par exemple en s'appuyant sur l'intuition, l'expérience ou des principes d'ingénierie -- évitant ainsi le vice de la correction des ``hypothèses après connaissance des résultats'' \cite{cockburn2018hark}. En effet, l'explication et la généralisation avec une théorie pourra parfaitement intervenir dans un second temps, dans des recherches ultérieures, par des méta-analyses, voire des expérimentations supplémentaires de vérification d'hypothèses.

\subsection{Faciliter la publication des réplications}

La pression de productivité pousse les recherches à focaliser sur le développement technologique, diminue les opportunités d'oser des hypothèses risquées, de trouver des résultats inattendus et d'obtenir des résultats inhabituels \cite{greenberg1992weak,murayama2015management,yaqub2018taxonomy}, et borne la recherche autour de quelques méthodes dominantes garantissant la publication \cite{greenberg2008usability,yaqub2018taxonomy}. Travailler depuis des cas applicatifs, optimisant ainsi divers prototypes et testant diverses tâches applicatives, devrait créer un cadre favorable à l'émergence de la sérendipité et permettre de convertir des résultats inattendus en construction de nouvelles théories, comme déjà éprouvé par certaines sciences des organisations \cite{eisenhardt1989building}. Ainsi, les cas applicatifs peuvent participer de plusieurs buts, depuis l'illustration d'applications pertinentes des technologies, la mise à l'épreuve de théories, aussi bien que leur émergence \cite{eisenhardt1989building}. Dès lors, le traitement de cas applicatifs, qui a tendance à plutôt recevoir des évaluations sévères \cite{greenberg2008usability}, devrait voir croître ses opportunités de publication.

La nécessité de varier les tâches, pour rendre les études expérimentales plus informatives, a elle aussi été exprimée \cite{hornbaek2015wrong}. Or, traiter des cas applicatifs est un bon moyen de répondre à cette attente. En effet, varier les tâches parmi des cas applicatifs (par des tâches de bas niveau et des tâches représentatives de haut niveau) et entre des cas applicatifs (par des tâches répondants à des cadres applicatifs variés), contribuera à répondre à ce besoin de réplication sur une variété des tâches. Somme toute, les études de réplication souffrent d'être publiées pour être peu estimées, peu réussies ou peu controversées \cite{greenberg2008usability,hornbaek2014once}. Nonobstant, permettre des études de réplication par le traitement de cas applicatifs permettra de valoriser leur publication pour de multiples autres raisons, comme une bonne analyse du cas applicatif, une bonne conception du prototype, un développement technologique innovant, des études testant la conception, et des études comparant les technologies. Finalement, les études de propriétés rendent les cas applicatifs plus informatifs et valorisables pour leur publication, augmentant -- dans un cercle vertueux -- la chance des résultats d'être répliqués. Cette fusion -- opérée grâce à ces études duales -- a donc tout le potentiel pour des publications solides, qui contribueront à combler le ``grand vide'' de l'IHM \cite{reeves2015locating,kostakos2015hole} en bâtissant une science des propriétés interactionnelles pertinentes, de sorte à étudier l'interaction plutôt que les interfaces \cite{mbl2004designing}.

\section{Directions pour de futures recherches}\label{sec-future-work} 

Les futures recherches exploreront et adapteront les méthodes de méta-analyses et d'induction adéquates pour servir les efforts d'intégration et de généralisation de propriétés interactionnelles, et s'appliqueront à définir les relations qui peuvent exister entre les propriétés interactionnelles, en procédant, par exemple, par diffractions de multiples prismes ou par prismes à diffraction multiple selon des propriétés interactionnelles déjà établies, voire en ébauchant une physique des interactions humain-machine. Ces recherches bénéficieront d'avoir mieux cartographié et normalisé les noms des boucles d'interaction impliquées dans les différents genres d'interfaces humain-machine, de sorte à pouvoir raffiner la description des conditions expérimentales et des propriétés interactionnelles.

En outre, la définition actuelle de propriété interactionnelle, ainsi que la méthode de révélation associée, tend à se concentrer sur les propriétés ergonomiques de l'interaction. Un prolongement de la présente recherche serait donc de considérer aussi l'interaction, par exemple, visant à faire vivre une expérience et de déterminer si l'altération de boucles, la réplication de résultats et la conduite de méta-analyses restent alors encore pertinentes.

\section{Conclusion}\label{sec-conclusion}

Nous proposons une nouvelle approche en IHM, pour aboutir à une science de propriétés interactionnelles, obtenues en multipliant les observations sur des cas applicatifs, dans un premier temps, puis par méta-analyses et, éventuellement, induction de théories, pour expliquer et généraliser les résultats, dans un deuxième temps. L'ambition est de répéter les mesures à l'échelle des communautés en IHM, plutôt que se contenter de mesures isolées \cite{greenberg1992weak,greenberg2008usability,hornbaek2014once,hornbaek2015wrong}, qui souvent s'appuient sur la signification statistique comme ultime preuve \cite{cockburn2020threats}, et surgénéralisent sans appuis théoriques \cite{greenberg1992weak}. Les cas applicatifs servent alors de moteur d'impulsion et de diversification des réplications.

Dans cette optique, le présent article introduit un cadre conceptuel permettant de caractériser la motivation des études expérimentales, de sorte à proposer de nouvelles études duales, optimisant l'utilisation de prototypes et chassant l'existence de causalités, appelées études de propriétés interactionnelles. L'objet de ces études est de révéler l'interaction en diffractant les boucles d'interaction en des points uniques, à l'aide de prismes interactionnels, un outil imaginaire à l'image des prismes utilisés en optique ondulatoire.

De la sorte, ces études permettent d'observer l'interaction sur des variétés de prototypes applicatifs et d'exprimer des résultats calibrés dans un même formalisme,
permettant ainsi d'accumuler des ensembles d'observations répliquées entre des variétés de prototypes, de technologies, de tâches et de profils de personnes utilisatrices. Ces observations participeront aussi bien à alimenter l'émergence de futures théories, que de mieux préciser les conditions dans lesquelles des théories existantes s'appliquent~: l'approche est donc fondée scientifiquement. Aussi, ces études, en s'appuyant sur des prototypes applicatifs, demandent peu d'efforts additionnels, tout en facilitant leur publication et en participant à leur optimisation~: l'approche est donc également fondée économiquement.

Nous envisageons que cette approche, illustrées par de premières études et de premiers cas applicatifs, puisse impulser une multiplication de résultats empiriques calibrés, puis que leur intégration et leur généralisation participera à combler le ``grand vide'' en IHM qui réside entre science et ingénierie \cite{kostakos2015hole,reeves2015locating}, par de nouvelles propriétés, lois et règles de conception pertinentes, et plus particulièrement hors du champ des interfaces universelles comme, par exemple, pour répondre à la pluralité de genres des interfaces de l'informatique ubiquitaire.

\selectlanguage{english}

\bibliographystyle{ACM-Reference-Format}
\bibliography{paper.bib}

\appendix

\end{document}